\documentclass[aps,prl,superscriptaddress,twocolumn]{revtex4}

\usepackage{graphicx}
\usepackage{dcolumn}
\usepackage{bm}

\usepackage[utf8]{inputenc}
\usepackage[T1]{fontenc}
\usepackage{etoolbox}
\usepackage{verbatim}
\usepackage{hyperref}
\usepackage{bbold}
\usepackage{xcolor}
\usepackage{mathrsfs}
\usepackage{amsmath,amsfonts,amssymb}
\usepackage{braket}
\usepackage{url}
\usepackage{subfigure}
\usepackage[dvipsnames]{xcolor}

\usepackage[english]{babel}

\newcommand{\beq}{\begin{equation}}
\newcommand{\eeq}{\end{equation}}
\newcommand{\beqa}{\begin{eqnarray}}
\newcommand{\eeqa}{\end{eqnarray}}

\begin{document}


\title{\textcolor{black}{Sparse SYK quantum batteries:} Boosting quantum efficiency by reducing complexity}

\author{Giovanni Sisorio}
\author{Alberto Cappellaro}

\affiliation{Dipartimento di Fisica ed Astronomia \textit{"G. Galilei"}, Universit\`{a} degli Studi di Padova, via Marzolo 8, 35131 Padova, Italy}

\author{Luca Dell'Anna}
\email{luca.dellanna@unipd.it}

\affiliation{Dipartimento di Fisica ed Astronomia \textit{"G. Galilei"}, Universit\`{a} degli Studi di Padova, via Marzolo 8, 35131 Padova, Italy}

\affiliation{National Institute of Nuclear Physics (INFN), Padova Section, via Marzolo 8, 35131 Padova, Italy}

\date{\today{}}

\begin{abstract}
In the context of energy storage at the nanoscale, exploring the notion of \textit{quantum
advantage} implies walking on the thin line at the boundary between quantum mechanics and
thermodynamics, which underpins our conventional understanding of battery devices. 
With no classical analogue, the Sachdev-Ye-Kitaev (SYK) model has emerged in the
last years as a promising platform to boost charging and storage efficiency thanks to 
its highly-entangling dynamics. Here, we explore how the robustness of this setup changes by considering
the sparse version of the SYK model, showing that, as long as chaos is not completely broken,
reducing its complexity may lead to even more efficient quantum batteries \textcolor{black}{amplifying the charging power}. 
\end{abstract}

\maketitle

\textit{Introduction and motivations. } 
The quest for efficient energy storage and work extraction has been a cornerstone of several technological breakthroughs, dating back at least to the first industrial revolution, with its profound connection to the development of thermodynamics as the backbone for our scientific understanding of Nature \cite{pierno-book, maxwell-demon-book}. Nowadays, as material samples are brought to increasingly lower temperatures and smaller sizes, we are experiencing a nanoscale industrial revolution. Here, the challenge of energy storage takes on a quantum mechanical dimension, where centuries-old thermodynamic concepts must be revisited in the light of non-classical resources such as quantum coherence and entanglement \cite{halpern-2022, kurizki-2022, strasberg-2024}. Indeed, the interplay between entanglement generation and work extraction from many-body quantum states has spurred a sustained research effort into quantum batteries \cite{alicki-2013, binder-2015, ferraro-2018, andolina-2018, liu-2019, rossini-2019, quach-2022, barzanjeh-2024, campaioli-2024}.

It is now clear that quantum battery engineering hinges on balancing complexity and controllability. For instance, while being highly sensitive to noise and fluctuations, the onset of highly entangling dynamics can positively affect battery performance, especially regarding figures of merit such as the amount of injected energy or the charging power \cite{hovhannisyan-2013, caravelli-2020, campaioli-2024}. This tension motivates the exploration of minimal platforms where access to such observables is amenable to analytical and computational analysis.

One of the most interesting candidates is the Sachdev-Ye-Kitaev (SYK) model, a paradigmatic example of a strongly interacting quantum system describing all-to-all randomly interacting fermions \cite{sachdev-1993, kitaev-2015-p1, kitaev-2015-p2, sachdev-2015, gu-2020}. This peculiar structure supports the onset of maximal chaos, such that the system qualifies as a fast scrambler, leading to very efficient charging and stable energy storage \cite{swingle-2020, rossini-2020, rosa-2020}. However, it also presents significant experimental and computational challenges. The former are a direct consequence of the all-to-all character of SYK couplings, which results in significant practical bottlenecks regarding system size \cite{danshita-2017, chen-2018, can-2019, hauke-2023, hauke-2024}. In addition, all-to-all interactions lead to volume-law entangled states, such that even powerful techniques like tensor networks become significantly costly \cite{bettaque-2024}. Consequently, exact diagonalization is limited to a few tens of fermions due to the rapid scaling of the Hilbert space size \cite{fu-2016, balents-2018, zhang-2022}.

In this work, we investigate the sparse version of the SYK model, where interaction terms are removed with a certain probability \cite{xu-2020, garcia-garcia-2021, preskill-2024}. By \textit{pruning} a fraction of the couplings, quantum chaotic features—instrumental for efficient energy storage—are retained while mitigating computational and experimental overhead. Even more interestingly, our results demonstrate that such sparsification can actually enhance battery performance, provided the system remains sufficiently chaotic.

\begin{figure}[htb!]
\centering
    \begin{subfigure}
    \centering
    \includegraphics[width=0.9\linewidth]{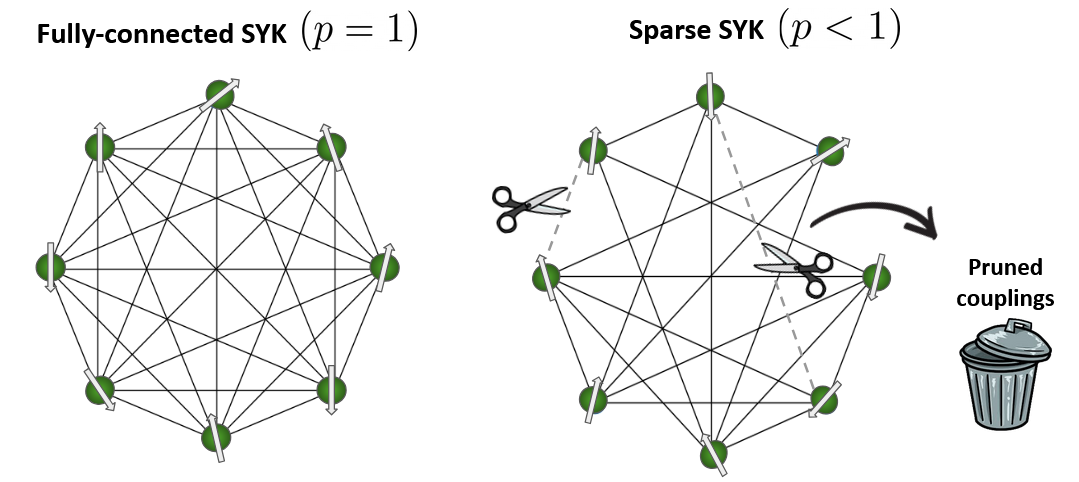}
    \end{subfigure}
    \begin{subfigure}
    \centering
    \includegraphics[width=0.9\linewidth]{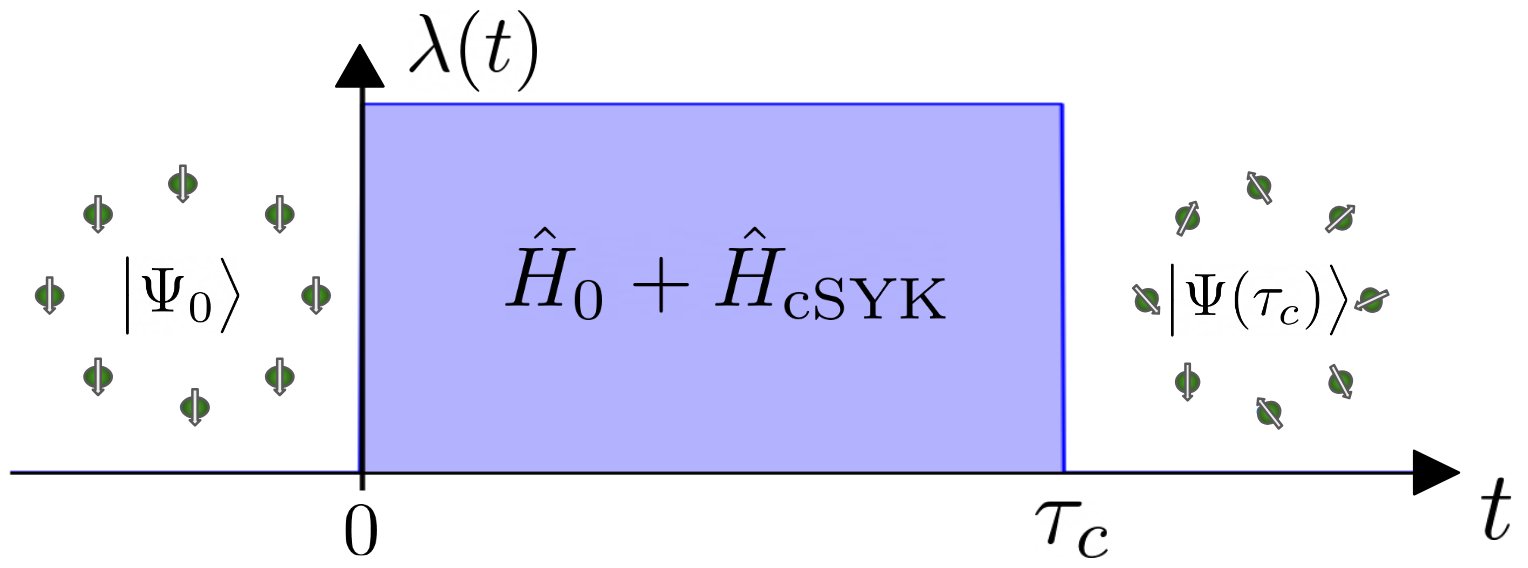}
    \end{subfigure}
\caption{\textit{Top panel.} Pictorial representation of the fully-connected SYK model, as opposed
to its sparse version. In the fully-connected version, 
we are considering a set of $N$ all-to-all randomly interacting (complex) fermions. 
Once we move to the sparse version, couplings are retained with probability $p$ (the so-called
sparsity parameter, which is $1$ in the fully-connected case), and otherwise
removed. Also notice that, in principle, the complex SYK model defined in Eq. \eqref{complex SYK}
accounts for spinless fermions, with the mapping to two-level systems (i.e. the battery
elementary cells) enabled by the Jordan-Wigner transformation \cite{SM}.
\textit{Bottom panel. } Charging protocol of a quantum battery in a nutshell, as described in the main text 
(cfr. Eq. \eqref{battery hamiltonian: charging protocol} and discussion thereafter). After
preparing our set of $N$ two-level systems in its ground state $\big|\Psi_0\big\rangle$, we turn
on the charging Hamiltonian $\hat{H}_{\text{cSYK}}$ at a certain value of the sparsity parameter,
and we let the dynamics unfold up to $\tau_c$. Accessing the evolved state 
$\big| \Psi(\tau_c)\big\rangle$ is paramount to unveil the quantum battery performance, through properly defined figures of merit.}
\label{fig:intro}
\end{figure}

\textit{The sparse SYK model. } 
As mentioned in the introduction, the SYK model presents itself as a theory for strongly
coupled fermions with random (all-to-all) interactions 
\cite{sachdev-1993,kitaev-2015-p1,kitaev-2015-p2}. Now, in recent years, most
of the existing efforts in elucidating the peculiar properties of this model focused on the
case with Majorana fermions. Here, on the other hand, 
our starting point is the complex version of the SYK (cSYK) model 
\cite{sachdev-2015,gu-2020} for $N$ spinless fermions, i.e.
\begin{equation}
\hat{H}_{\text{cSYK}} = \sum_{i,j,k,l = 1}^N J_{ijkl} \;\hat{c}^{\dagger}_i
\hat{c}^{\dagger}_j\hat{c}_k\hat{c}_l 
\label{complex SYK}
\end{equation}
with the usual anticommutation relations 
$\lbrace \hat{c}_i,\hat{c}^{\dagger}_j \rbrace = \delta_{ij}$ and
$\lbrace \hat{c}^{(\dagger)}_i,\hat{c}^{(\dagger)}_j \rbrace = 0$, while the 
complex couplings satisfy $J_{ijkl} = J^*_{klij}$ and $J_{ijkl} = - J_{jikl} = - J_{ijlk}$,
such that Eq. \eqref{complex SYK} is Hermitian and anti-symmetrized. These couplings are distributed according to a zero-mean Gaussian density function $\mathcal{P}(J_{ijkl})$, 
with variance  $\overline{J^2_{ijkl}} = J^2/N^3$, the average over disorder
being defined as $\overline{\mathcal{O}(J_{ijkl})} = \int \mathcal{D}[J_{ijkl}] \mathcal{P}(J_{ijkl})
O(J_{ijkl})$. This model can conveniently be mapped to a spin model by the Jordan-Wigner transformation (see \cite{SM}).  
The sparse version of Eq. \eqref{complex SYK} is simply implemented by introducing, 
for each coupling, 
an additional random variable $x_{ijkl}$ being equal to $1$ with probability $p$ and $0$
otherwise, such that the coupling is retained with probability $p$ or \textit{pruned} 
(i.e. removed) with probability $1-p$ \cite{xu-2020,garcia-garcia-2021}.
Therefore, for a $q$-body interaction 
(in the case of Eq. \eqref{complex SYK}, $q = 4$), the \textit{pruning} procedure 
reduces the number of interacting terms $\binom{N}{q}$ by a factor $p$. In order to ensure
comparability with the energy scale of the original model (i.e. $p = 1$), 
$\overline{J^2_{ijkl}}\longrightarrow \overline{J^2_{ijkl}} / p$. 
%
%

Even in its complex version, it is important to recall that the 
SYK model is maximally chaotic, saturating a general bound \cite{maldacena-2016} 
on the Lyapunov exponent $\lambda_L$ which controls the decay of out-of-time four-point correlators
as $\big\langle \hat{A}(t) \hat{B}(0) \hat{A}(t) \hat{B}(0) \big\rangle_{\text{th}} 
\sim 1 - \alpha \exp (\lambda_L t_s)$, with $\hat{A}$ and $\hat{B}$ some system's operators,
$\langle \bullet \rangle_{\text{th}}$ the thermal average and $t_s$ the so-called scrambling time.
The onset of maximal chaos implies that information is scrambled at the fastest rate enabled
by quantum mechanics and it is crucial for an efficient charging protocol 
\cite{rossini-2020,rosa-2020,swingle-2020}. 
It is then natural to investigate the 
relation between sparsity and quantum chaos, especially when one has to leverage the latter
to implement an efficient setup. Technically, the quantum chaotical character of the SYK model
reflects on specific features of its energy spectrum, which can be analyzed by means of tools 
from random matrix theory (RMT) \cite{guhr-1998,dalessio-2016,ludwig-2017,preskill-2024}. 
Here, we first examine the behavior of the nearest-neighbor gap ratio, which is defined as 
\begin{equation}
r = \bigg\langle\min \bigg(\frac{s_i}{s_{i+1}}, \; \frac{s_{i+1}}{s_i}\bigg) \bigg\rangle\;,
\label{gap ratio definition}
\end{equation}
with $s_i = E_i - E_{i-1}$ the spacing between adjacent eigenvalues. The gap ration
quantifies the repulsion between energy levels at scales of the mean level spacing. 
Consistently with predictions from RMT, $r$ remains approximately constant for large 
$p$, while a sharp drop is observed as soon as $p$ approaches a critical value 
of sparsity labelled $p_c$, signalling the dramatic breakdown of spectral rigidity.
\begin{figure}[h!]
    \centering
    \includegraphics[width=0.9\linewidth]{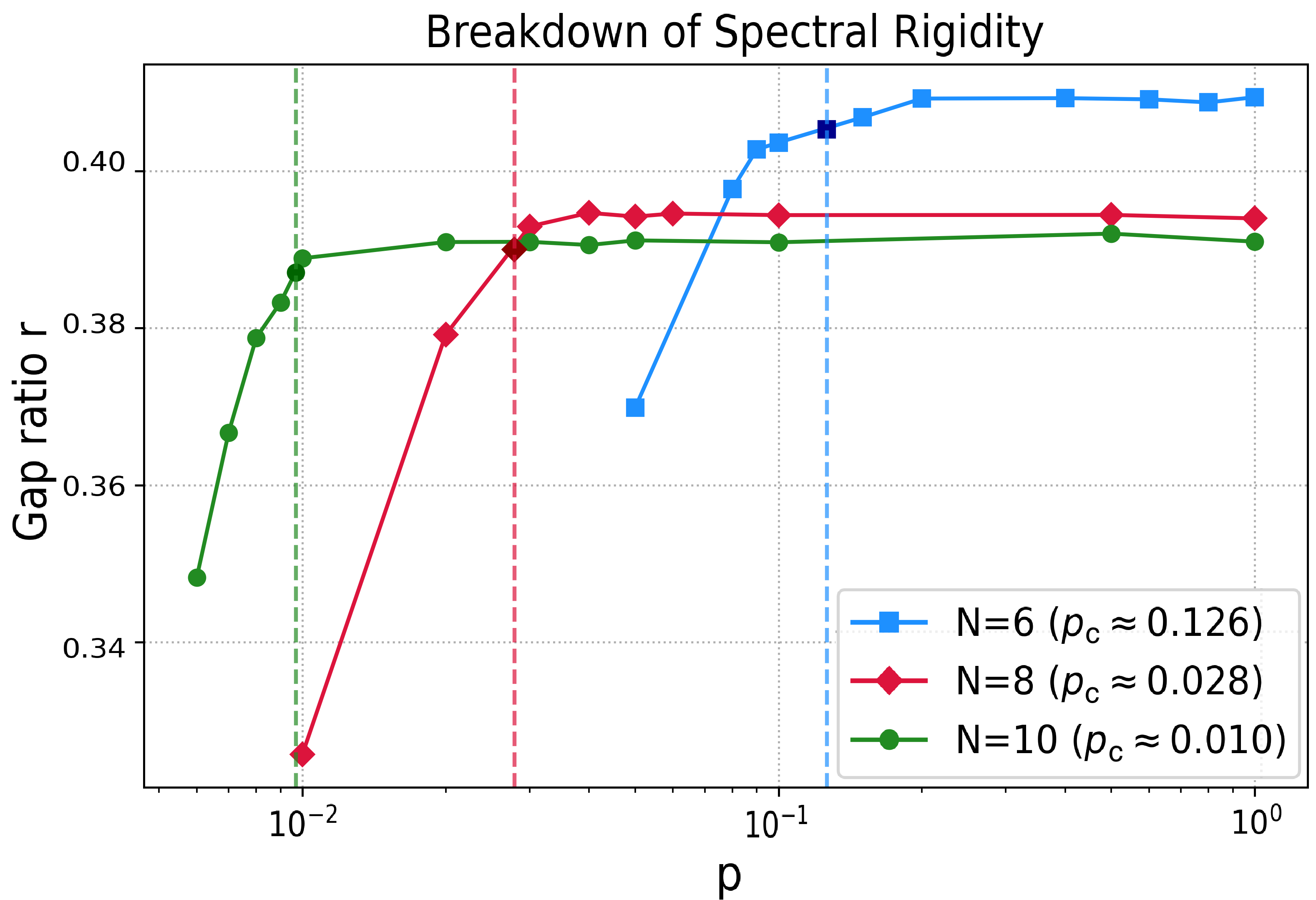}
    \caption{Nearest-neighbor eigenvalue gap ratio $r$ (see. Eq. \eqref{gap ratio definition}) 
    as a function of the sparsity parameter $p$ for $N = 6$ (cyan squares), $N = 8$
    (red diamonds), and $N = 10$ (green circles), with $J = 1$. For large $p$, $r$ remains approximately constant, consistent with the Random Matrix Theory (RMT) predictions. As $p$ decreases, a sharp decline in $r$ indicates the breakdown of spectral rigidity and the transition to 
    non-universal statistics. 
    The critical sparsity values $p_c$ are highlighted in a darker shade: $p_c = 0.1263$ for $N = 6$, $p_c = 0.0279$ for $N = 8$, and $p_c = 0.0097$ for $N = 10$. Statistical errors originating from
    the disorder average are included, but they end up smaller than the marker size. 
    }
    \label{fig: p2_values}
\end{figure}
%

In Fig. \ref{fig: p2_values} we report the gap ration $r$ as in 
Eq. \eqref{gap ratio definition} as a function of the sparsity parameter $p$ 
and for three different values of $N$, clearly displaying the behavior we have just
outlined.
From an operational perspective, we define the critical sparsity value $p_c$ as the value of $p$ 
where $r$ drops below $99\%$ of its value in the fully connected (unsparsified) case
\cite{preskill-2024}. At this threshold, the characteristic \textit{ramp} 
in the spectral statistics vanishes, marking a qualitative change in the eigenvalue 
dynamics.  \textcolor{black}{The critical sparsity parameter $p_c$ approaches zero upon increasing $N$ consistently with   
\begin{equation}
p_c=\frac{k_c N}{N_J}, 
\end{equation}
where $N_J$ is the number of interaction terms, which, in our case is $N_J=
 \binom{N}{2}^2$, and $k_c\approx 2.4$ (see \cite{SM} for details). For very large $N$ it vanishes as ${4 k_c}/{N^3}$. A similar behavior is reported in Refs.~\cite{xu-2020, preskill-2024} with a different $N_J$, namely $N_J=\binom{N}{4}$, since there SYK model is composed by Majorana fermions, but with nearly the same critical $k_c$, i.e. the same number of remaining hyperedges $k_cN$.}
%
%
%

\textit{Implementing the battery: The charging protocol. }
A quantum battery can simply be intended 
as a collection of $N$ quantum cells \cite{alicki-2013,binder-2015,campaioli-2024}, 
each acting as an individual unit of energy
storage such that $\hat{H}_0 = \sum_{i=1}^N \hat{h}_i$. Here, we focus on
the charging protocol, thus assuming that the battery is initially prepared 
in its discharged state
(the lowest eigenstate of $\hat{H}_0$) up to $t\rightarrow 0^-$. Then, at $t = 0$ 
the charging Hamiltonian is turned on, realizing the following protocol
\begin{equation}
\hat{H}_B(t) = \hat{H}_0 + \lambda(t) \hat{H}_1
\label{battery hamiltonian: charging protocol}
\end{equation}
where $\lambda(t) = 1$ when $0 \leq t < \tau_c$, and $0$ elsewhere, with $\tau_c$ 
being the duration of the charging protocol. 
\textcolor{black}{Differently from \cite{rossini-2020}, 
we consider a charging protocol governed by the full Hamiltonian (\ref{battery hamiltonian: charging protocol}), as in Ref. (\cite{rosa-2020}). This choice is motivated by experimental considerations, as the internal energy of the battery cells, represented by $\hat{H}_0$, remains present during the charging phase.}
Now, within our implementation, $\hat{H}_0$
is an ensemble of $N$ two-level systems (or spin-$1/2$ particles, in the context of the JW transformation defined in \cite{SM}), 
while the sparse cSYK is responsible for charging. With the convention $\hbar = 1$,   
\begin{equation}
\hat{H}_0 = \omega_0 \sum_{i=1}^N \sigma^y_i\qquad
\text{and} 
\qquad
\hat{H}_1 = \hat{H}^{\text{(sparse)}}_{\text{cSYK}}
\label{sparse battery}
\end{equation}
where $\omega_0$ is the level splitting and $\sigma_y^i$ the usual Pauli matrix
for the $i^{\text{th}}$. 
%
%
\begin{figure}[ht!]
    \centering
    \includegraphics[width=0.93\linewidth]{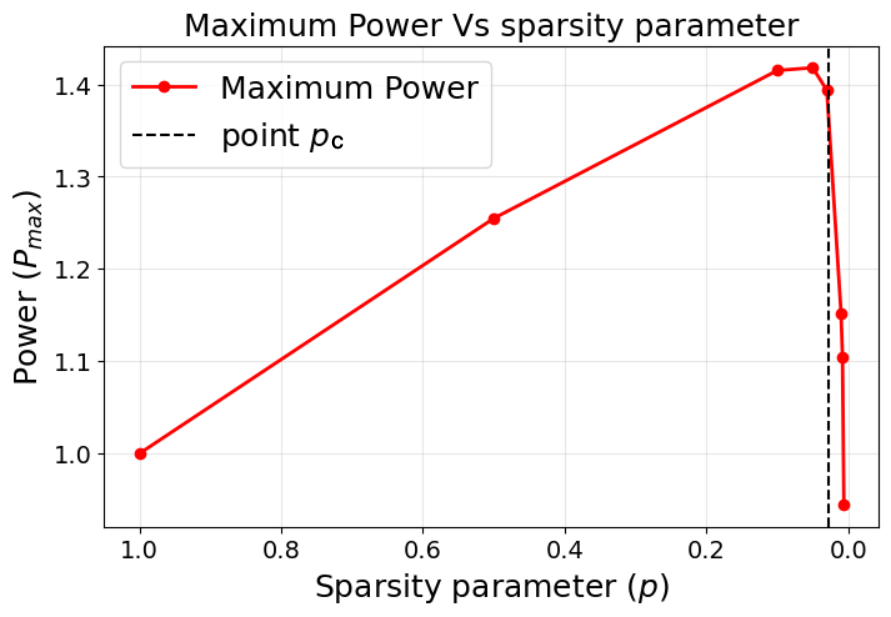}
    \caption{Maximum charging power $P_{\max}$ as a function of the sparsity parameter $p$ for $N=8$. As predicted, the power reaches its peak near the critical value $p \approx p_c$, where the system \textcolor{black}{ceases to exhibit the spectral signatures of quantum chaos}. 
    In the limit $p \to 0$, the power vanishes as the charging Hamiltonian $H_1$ is completely annihilated. 
    }
    \label{fig: stored energy}
\end{figure}
%
\textcolor{black}{To assess the performance of the sparse SYK         battery throughout the charging process, it is crucial to       examine specific figures of merit, with particular              attention to the role played by the sparsity parameter, like the stored energy or the population dynamics (see \cite{SM}). 
A       key quantity in this context is the average charging power,     defined as $P(\tau_c) = E_N(\tau_c)/\tau_c$, where the          stored energy $E_N(\tau_c)$ is given by:
    \begin{equation}
    E_N(\tau_c) = \big\langle \Psi(\tau_c) \big| \hat{H}_0 \big| \Psi(\tau_c)\big\rangle - E_N^{(0)}.
    \label{stored_energy}
    \end{equation}
    Here, $E_N^{(0)} = \big\langle \Psi_0|\hat{H}_0| \Psi_0\big\rangle$ represents the energy of the initial state $\big| \Psi_0 \big \rangle = \bigotimes^N_{i=1} \big|\downarrow^{(y)}\big\rangle_i$. 
    In Fig.~\ref{fig: stored energy}, we report the maximum charging power, $P_{\max} = \max_{\tau_c} P(\tau_c)$, as a function of the sparsity parameter $p$ for $N=8$. Remarkably, by reducing $p$ (i.e., increasing sparsity), we observe a significant enhancement in the maximum power, which reaches its peak value near the critical threshold $p \approx p_c$. This suggests a non-trivial trade-off between the onset of quantum chaos (and the resulting fast-scrambling dynamics) and the complexity emerging from an all-to-all interacting implementation. We found that $P_{\max}$ is improved by up to $40\%$ by pruning the connectivity of the model. 
    The fact that $P_{\max}$ reaches an optimal value around $p_c$ provides a first hint that battery performance is based on retaining at least some degree of the quantum-chaotic behavior characteristic of the fully connected ($p = 1$) case. Furthermore, for very small values of $p$, the power vanishes as expected, since the charging Hamiltonian $\hat{H}_1$ is essentially annihilated. These results indicate that pruning the interaction links not only facilitates experimental achievability but also represents an ideal configuration for maximizing the charging rate.}

\begin{figure}
    \centering
    \begin{subfigure}
    \centering
    \includegraphics[width=0.9\linewidth, height=0.9\linewidth]{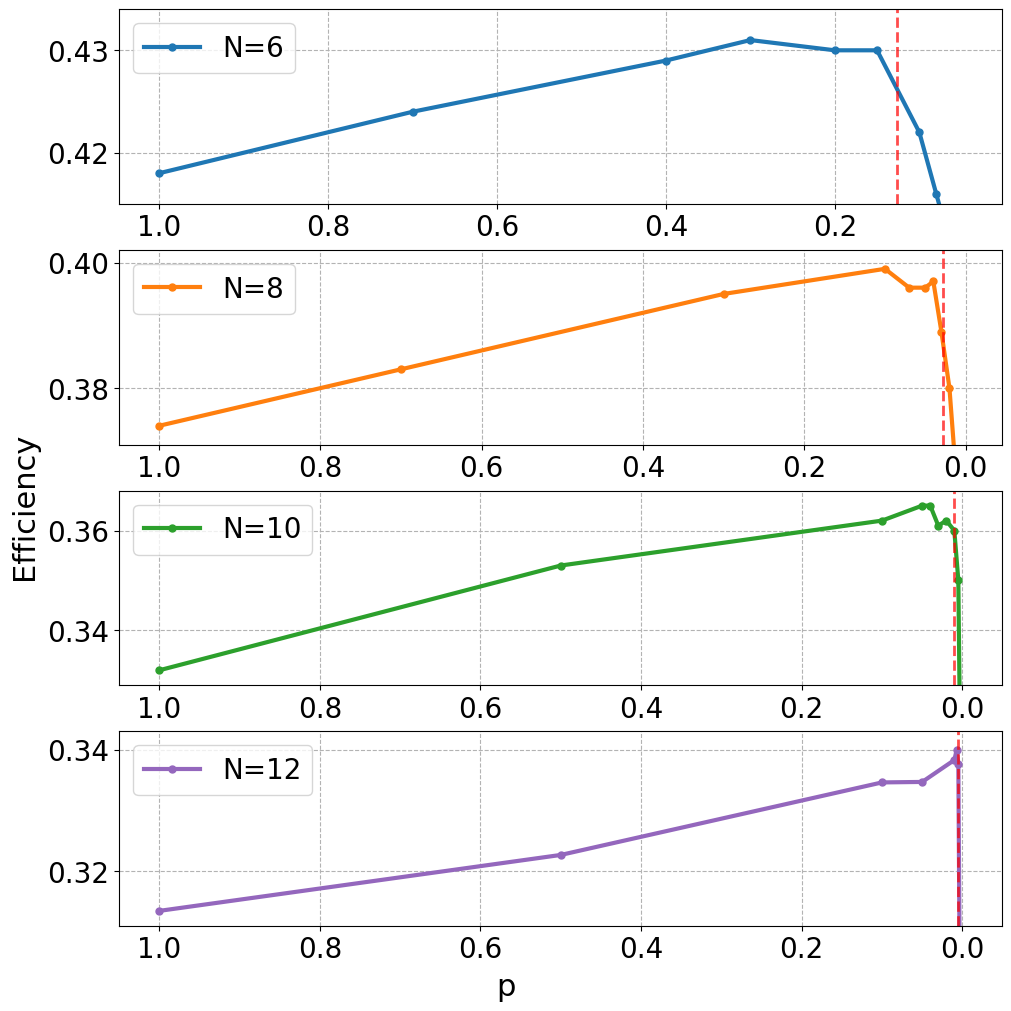}
    \end{subfigure}
    \caption{
    Battery efficiency, as defined in Eq. \eqref{battery efficiency}, with respect
    to different values of the sparsity parameter $p$, for three different system's size. 
    Here, we consider $\omega_0 = 1$, $J = 4.5$,  
    and the disorder average is performed 
    over $N_{\text{dis}} = 10^3$ for $N = 6$, $N_{\text{dis}} = 5\cdot 10^2$ for $N = 8$, $N_{\text{dis}} = 150$ for $N = 10$ and $N_{\text{dis}} = 100$ for $N=12$. Statistical uncertainties arising from 
    the disorder average are smaller than marker dots. 
    In the fully connected regime ($p = 1$), 
    it is evident how smaller batteries provides
    a more efficient avenue for work extraction. In the intermediate regime, moderate sparsity slightly improves performance. As $p$ approaches the critical value $p_c$ (dashed vertical lines), the breakdown of spectral rigidity and loss of chaotic features lead to a sharp decline in the efficiency.
    }
    \label{fig: E_vs_p}
\end{figure}
A more complete insight into this trade-off can be acquired by investigating 
a different figure of merit, ideally encoding additional information concerning
the work extraction at later times $t > \tau_c$ (i.e. after the charging protocol is 
turned off). This is the case for the so-called battery \textit{efficiency} \cite{rossini-2019},
defined as the ratio between the amount of extractable work and the energy stored
in the battery during the charging process. Technically, this is computed as 
\begin{equation}
e(\tau_c) = \overline{\bigg(\frac{\mathcal{E}_{N/2}(\tau_c)}{E_{N/2}(\tau_c)}\bigg)}
\label{battery efficiency}
\end{equation}
with $E_{N/2}(\tau_c)$ the stored energy as in Eq. \eqref{fig: stored energy} and
$\mathcal{E}_{N/2}(\tau_c)$ the maximum extractable work, also called \textit{ergotropy}
\cite{allahverdyan-2004,tirone-2021,touil-2021,campaioli-2024}. The latter is defined by
\begin{equation}
\mathcal{E}(t) \equiv \mathcal{E}(\rho(t)) = \text{Tr}\big[\hat{H}_0 \rho(t)\big]
- \text{Tr}\big[\hat{H}_0 \sigma_{\rho}\big]\;,
\label{ergotropy: definition}
\end{equation}
with $\sigma_{\rho}$ being a completely passive state (thermal states are good candidates
\cite{lenard-1978}), and $\rho(t)$ the state evolved according to the charging Hamiltonian, cfr. 
Eq. \eqref{battery hamiltonian: charging protocol}. Also notice that
in Eq. \eqref{battery efficiency} we are considering the stored energy (and the corresponding
extractable work) in half of the battery. Assuming that $\hat{H}_0$ can be recast as a sum
of local terms (this is evident from our setup Eq. \eqref{sparse battery}), we just need to 
restrict $\hat{H}_0$ in defining $\mathcal{E}(t)$ to the desired subset of elementary cells
(i.e. we stop the sum at $N/2$). Technically, this restriction reflects the reasonable assumption
that, in a realistic experimental platform, only a subset of the total number of qubits can
be accessed. 
In Fig. \ref{fig: E_vs_p} we report our numerical results concerning the battery efficiency
as defined in Eq. \eqref{battery efficiency} for different values of the sparsity parameter, with different comments now in order. First of all, even in the fully connected case
($p = 1$) increasing the battery size worsens its efficiency, a feature persisting even for 
different sparse realizations. 
This is in line with the seminal observation made in \cite{rosa-2020}, where it was first shown how
energy extraction from large SYK batteries is less favourable than dealing firsthand with
smaller batteries \textcolor{black}{(see \cite{SM})}.
Now, as sparsity gradually increases some interactions are removed, thus reducing complexity and,
at the same time, crucially retaining the quantum chaotic character typical of the fully-connected
regime. We discussed this previously, in relation to important markers of quantum chaos such as the 
nearest-neighbour ratio (see Fig. \ref{fig: p2_values}) or the spectral form factor dynamics (see \cite{SM}). 
Here the crucial point is that, in this
intermediate regime of reduced complexity, the battery appears to operate more efficiently 
likely because of the reduced interference between the interaction terms. 
This is evident by looking, for instance, at the $N = 10$ case 
(bottom panel in Fig. \ref{fig: E_vs_p}) where, by approaching $p_c$
from above, efficiency can be boosted by $\sim 10\%$. Remarkably, 
while bigger cSYK batteries are in principle less convenient for energy extraction, 
at the same time they seem to be 
more positively affected by an increased sparsity, when compared to smaller implementations
($N = 6$ and $N = 8$ in Fig. \ref{fig: E_vs_p}).
Finally, as we approach the critical value $p_c$, spectral rigidity breaks down and 
eigenvalue statistic deviate markedly from RMT predictions, signaling the loss of quantum chaos
and, consequently, the system's ability to efficiently scramble information. Therefore, 
the battery performance severely deteriorates in the regime $p < p_c$, as confirmed by 
our numerical results in Fig. \ref{fig: E_vs_p}.

\textcolor{black}{
    To further characterize the performance of the sparse SYK battery and quantify its charging agility, we analyze the scaling of the average quadratic energy fluctuations as a function of $N$.
    Specifically, we define the following quantity (\cite{rossini-2020}) 
    \begin{equation}
        \Delta_\tau \hat{H}^2_\alpha \equiv \frac{1}{\tau} \int_0^\tau \mathrm{d}t \left[ \langle \hat{H}^2_\alpha \rangle_t - \langle \hat{H}_\alpha \rangle_t^2 \right] \label{average_quadratic},
    \end{equation}
    where the index $\alpha \in \{0, 1\}$ refers to the internal Hamiltonian (connected with the distance traveled in the Hilbert space) and the interaction/charging Hamiltonian (representing the charging speed in the Hilbert space), respectively. 
    This quantity serves as a key figure of merit to quantify the system's capacity for coherent energy storage and the overall agility of the charging process and gives the following bound for the charging power 
    $P(\tau)\le 2\sqrt{\Delta_\tau \hat{H}^2_0 \, \Delta_\tau \hat{H}^2_1}$ \cite{julia-2020}.}
    
\begin{figure}
    	\includegraphics[height=4.3cm,width=4.3cm]{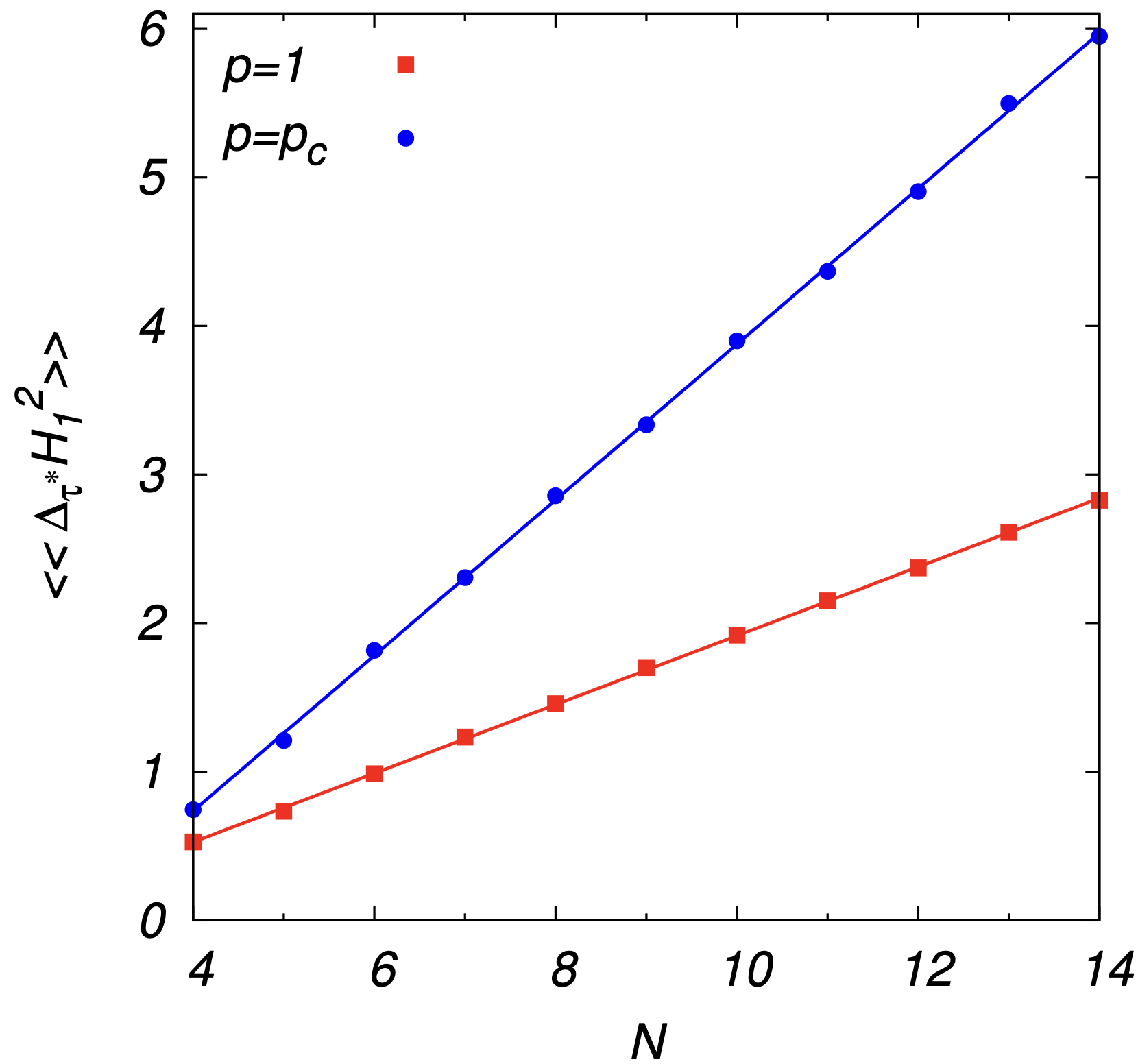}
        \includegraphics[height=4.3cm,width=4.33cm]{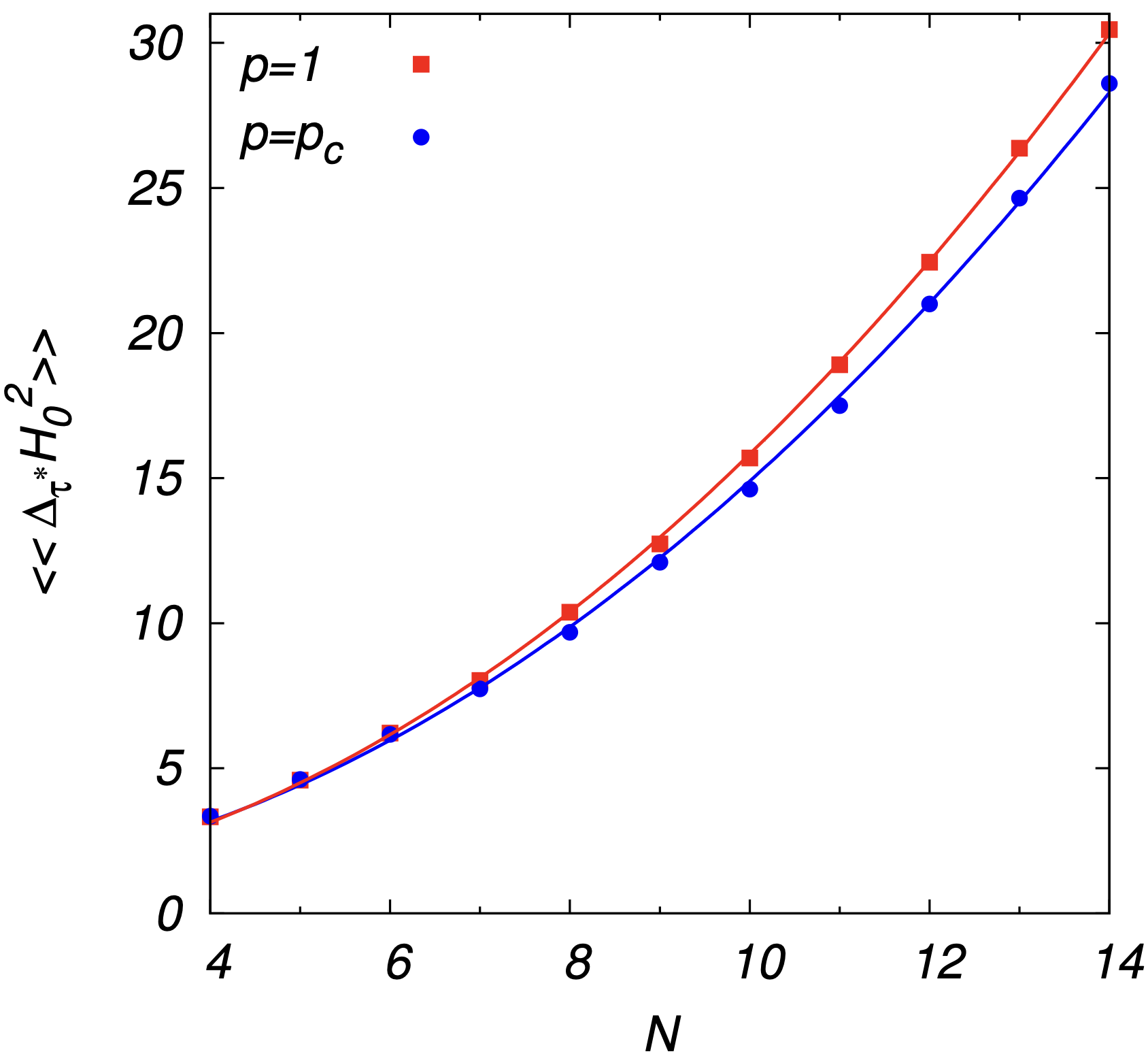}
    \caption{
    \textcolor{black}{
    Scaling with the system size $N$ of the quantities 
    $\langle\langle \Delta_{\tau^\star} \hat H_1^2 \rangle\rangle$, in units of $J^2$ (left figure), and $\langle\langle \Delta_{\tau^\star} \hat H_0^2 \rangle\rangle$, in units of $\omega_0^2$ (right figure), for two different sparsity regimes: 
    the red squares correspond to the fully connected case ($p=1$), while the blue circles refer to the sparse regime ($p=p_c$).
    Each point is obtained after averaging over nearly $200$ realizations of disorder. 
    The solid lines are the best fits with linear ($\sim N$, left panel) and quadratic ($\sim N^2$, right panel) functions. 
    } }
    \label{fig:Advantage_HO}
\end{figure}

  \textcolor{black}{  
    The results displayed in Fig.~\ref{fig:Advantage_HO} illustrate the scaling of $\langle\langle \Delta_{\tau^\star} \hat H_{\alpha}^2 \rangle\rangle$, namely $\Delta_{\tau^\star} \hat H_{\alpha}^2$ averaged over disorder \cite{rossini-2020}, with $N$, evaluated at the optimal time $\tau^\star$, for which $P(\tau^\star)=P_{max}$, for two distinct connectivity regimes: the fully connected case ($p=1$) and the sparse regime ($p=p_c$).
    The left panel shows the scaling behavior related to $\hat H_1$, \textcolor{black}{in units of $J^2$}. In the fully connected limit ($p=1$, red squares), the observable exhibits \textcolor{black}{a clear linear growth with $N$, whose best fit with the function $a N+b$ gives $a\approx 0.23$ and $b\approx -0.40$. 
    This behavior is characteristic of architectures where energy scales extensively with the system size. 
    In the sparse regime ($p=p_c$, blue circles), this quantity grows significantly faster, although linearly, with a slop which is greater  than twice that obtained for the fully connected case. The best fit with the a linear function $a' N+b'$ gives, indeed, $a'\approx 0.52$ and $b'\approx -1.36$.}
    Since this boost is driven solely by the tuning of the sparsity parameter $p$, it is identified as a \textit{non-genuine} quantum advantage, arising from the structural reconfiguration of the interactions rather than the intrinsic nature of the operators.
        The scenario changes when considering $\langle\langle \Delta_{\tau^\star} \hat H_0^2 \rangle\rangle$, \textcolor{black}{in units of $\omega_0^2$, as shown in the right panel of Fig.~\ref{fig:Advantage_HO}. Here, in the fully connected case ($p=1$), the observable displays a clear super-linear quadratic growth with $N$. The best fit with the function $c N^2+d$ is obtained for $c\approx 0.15$ and $d\approx 0.72$.}
    This trend, $\sim N^2$, is in agreement with the results reported in Ref.~\cite{rossini-2020} and represents the signature of a \textit{genuine} quantum advantage, intrinsically linked to the non-linear structure of $\hat H_0$. 
    \textcolor{black}{The introduction of sparsity ($p=p_c$) further confirm this effect, leading to an almost equally fast growth as in the case with $p=1$, thereby validating the quadratic behavior observed for the energy fluctuations related to the Hamiltonian $\hat H_0$. Also in this sparse case, fitting the numerical results with $c' N^2+d'$ we get $c'\approx 0.14$ and $d'\approx 0.94$.
    The results confirm that the quantum advantage is robust with respect to the specific charging protocol employed and the charging power is even reinforced. 
    Since, for any $J$ we have  $\langle\langle\Delta_{\tau^*} \hat{H}^2_1\rangle\rangle (p=p_c)\simeq 2 \langle\langle\Delta_{\tau^*} \hat{H}^2_1\rangle\rangle (p=1)$ while, for $J\gtrsim 3$, $\langle\langle\Delta_{\tau^*} \hat{H}^2_0\rangle\rangle (p=p_c)\simeq  \langle\langle\Delta_{\tau^*} \hat{H}^2_0 \rangle\rangle(p=1)$, we expect to have, for large $N$ (see \cite{SM} for more details)
    \begin{equation}
    P_{max}(p=p_c)\approx  \sqrt{2} \,P_{max}(p=1) \lesssim  \alpha J\omega_0 N^{3/2}
    \end{equation}
    with $\alpha\approx 0.5$, namely an enhancement of the charging power at the critical point by approximately $40\%$ with respect to the fully connected case.}
    This finding suggests that the observed enhancement is a fundamental consequence of the underlying interaction structure rather than a protocol-dependent artefact. 
    These findings provide a strong evidence that sparse charging Hamiltonians can serve as a powerful resource to overcome  the limitations of the fully connected architectures by combining genuine and structural quantum advantages.}


\textit{Experimental considerations.} 
It is certainly relevant that all the crucial features of the fully connected SYK model persist in
its sparse version, especially (at least for our goals) when it comes to the onset of maximal chaos.
Indeed, implementing a sparse model may enable experimentalists to overcome significant bottlenecks
displayed by the full version. Even within the framework of ultracold atomic gases,
where almost perfect isolation from the external environment and control on many physical parameters
are at hand, implementing the fully connected SYK would require $4\,$$\cdot$$10^3$ lasers just for
the $N = 16$ fermions \cite{danshita-2017}. A similar $N$ is achieved by considering mesoscopic graphene
flakes at strong disorder and magnetic field, by considering the electrons in the lowest Landau level
\cite{chen-2018}. Here, the bottleneck is represented by the magnetic field strength, which can reach 
$B \sim 3\cdot 10^3$ T for a flake with $5$ nm radius ($2\cdot 10^3$ carbon atoms mapping to the $N\lesssim 20$ scenario). While larger flakes enable more realistic conditions \cite{kruchkov-2023},  
implementing a sparse version represents an appealing alternative. This is even more compelling 
when considering the recent advances with novel platforms based on 
cavity quantum electrodynamical simulations \cite{hauke-2023,hauke-2024}. 

\textit{Conclusions. } Our work underscores the potential of sparse quantum systems as a viable
path toward the goal of implementing robust quantum batteries, offering a blueprint for
harnessing quantum resources without succumbing to impractical complexity. More specifically,
we have first shown that, starting from the fully connected SYK model, quantum chaotic features 
persist down to a critical value ($p_c$) of the sparsity parameter, a threshold 
where spectral rigidity breaks down. 
Remarkably, we show that 
moving closer to $p_c$ from above 
positively affects the battery 
performance.  Ultimately, this approach also sheds light on the fundamental interplay
between complexity, chaos and energy scale at the nanoscale.

\textit{Acknowledgements. } The authors acknowledge financial support from the Project PARD 2024 “Role of disorder in work extraction and energy storage for Sachdev-Ye-Kitaev quantum batteries” within the Project ”Frontiere Quantistiche” (Dipartimenti di Eccellenza) of the Italian Ministry for Universities and Research, and from the European Union-Next Generation EU within the “National Center for HPC, Big Data and Quantum Computing” (Project No. CN00000013, CN1 Spoke 10 - Quantum Computing). The authors thank M. Di Liberto for the concession of Virtual Machine at CloudVeneto, and F. Campaioli for stimulating discussion.
\textit{Code availability.} All the codes employed in this paper publicly available \cite{sisorio-2025}. 
They have been developed thanks to the open-source Python framework QuTiP 
\cite{johansson-2012,johansson-2013,lambert-2024-qutip5}.
%
%
\bibliography{references}

\providecommand{\noopsort}[1]{}\providecommand{\singleletter}[1]{#1}%
\begin{thebibliography}{60}
\expandafter\ifx\csname natexlab\endcsname\relax\def\natexlab#1{#1}\fi
\expandafter\ifx\csname bibnamefont\endcsname\relax
  \def\bibnamefont#1{#1}\fi
\expandafter\ifx\csname bibfnamefont\endcsname\relax
  \def\bibfnamefont#1{#1}\fi
\expandafter\ifx\csname citenamefont\endcsname\relax
  \def\citenamefont#1{#1}\fi
\expandafter\ifx\csname url\endcsname\relax
  \def\url#1{\texttt{#1}}\fi
\expandafter\ifx\csname urlprefix\endcsname\relax\def\urlprefix{URL }\fi
\providecommand{\bibinfo}[2]{#2}
\providecommand{\eprint}[2][]{\url{#2}}

\bibitem[{\citenamefont{Saggion et~al.}(2019)\citenamefont{Saggion, Faraldo,
  and Pierno}}]{pierno-book}
\bibinfo{author}{\bibfnamefont{A.}~\bibnamefont{Saggion}},
  \bibinfo{author}{\bibfnamefont{R.}~\bibnamefont{Faraldo}}, \bibnamefont{and}
  \bibinfo{author}{\bibfnamefont{M.}~\bibnamefont{Pierno}},
  \emph{\bibinfo{title}{Thermodynamics: Fundamental Principles and
  Applications}}, UNITEXT for Physics (\bibinfo{publisher}{Springer
  International Publishing}, \bibinfo{year}{2019}), ISBN
  \bibinfo{isbn}{9783030269760}.

\bibitem[{\citenamefont{Leff and Rex}(1990)}]{maxwell-demon-book}
\bibinfo{editor}{\bibfnamefont{H.~S.} \bibnamefont{Leff}} \bibnamefont{and}
  \bibinfo{editor}{\bibfnamefont{A.~F.} \bibnamefont{Rex}}, eds.,
  \emph{\bibinfo{title}{Maxwell's Demon: Entropy, Information, Computing}}
  (\bibinfo{publisher}{Princeton University Press}, \bibinfo{year}{1990}),
  \urlprefix\url{http://www.jstor.org/stable/j.ctt7zts1p}.

\bibitem[{\citenamefont{Halpern}(2022)}]{halpern-2022}
\bibinfo{author}{\bibfnamefont{N.}~\bibnamefont{Halpern}},
  \emph{\bibinfo{title}{Quantum Steampunk: The Physics of Yesterday's
  Tomorrow}} (\bibinfo{publisher}{Johns Hopkins University Press},
  \bibinfo{year}{2022}), ISBN \bibinfo{isbn}{9781421443737},
  \urlprefix\url{https://books.google.it/books?id=W_pSEAAAQBAJ}.

\bibitem[{\citenamefont{Kurizki and Kofman}(2022)}]{kurizki-2022}
\bibinfo{author}{\bibfnamefont{G.}~\bibnamefont{Kurizki}} \bibnamefont{and}
  \bibinfo{author}{\bibfnamefont{A.}~\bibnamefont{Kofman}},
  \emph{\bibinfo{title}{Thermodynamics and Control of Open Quantum Systems}}
  (\bibinfo{publisher}{Cambridge University Press}, \bibinfo{year}{2022}), ISBN
  \bibinfo{isbn}{9781316814574},
  \urlprefix\url{https://books.google.it/books?id=xdtXEAAAQBAJ}.

\bibitem[{\citenamefont{Strasberg}(2024)}]{strasberg-2024}
\bibinfo{author}{\bibfnamefont{P.}~\bibnamefont{Strasberg}},
  \emph{\bibinfo{title}{Quantum Stochastic Thermodynamics: Foundations and
  Selected Applications}}, Oxford Graduate Texts (\bibinfo{publisher}{Oxford
  University Press}, \bibinfo{year}{2024}), ISBN \bibinfo{isbn}{9780198931584},
  \urlprefix\url{https://books.google.it/books?id=gzOt0AEACAAJ}.

\bibitem[{\citenamefont{Alicki and Fannes}(2013)}]{alicki-2013}
\bibinfo{author}{\bibfnamefont{R.}~\bibnamefont{Alicki}} \bibnamefont{and}
  \bibinfo{author}{\bibfnamefont{M.}~\bibnamefont{Fannes}},
  \bibinfo{journal}{Phys. Rev. E} \textbf{\bibinfo{volume}{87}},
  \bibinfo{pages}{042123} (\bibinfo{year}{2013}).

\bibitem[{\citenamefont{Binder et~al.}(2015)\citenamefont{Binder,
  Vinjanampathy, Modi, and Goold}}]{binder-2015}
\bibinfo{author}{\bibfnamefont{F.~C.} \bibnamefont{Binder}},
  \bibinfo{author}{\bibfnamefont{S.}~\bibnamefont{Vinjanampathy}},
  \bibinfo{author}{\bibfnamefont{K.}~\bibnamefont{Modi}}, \bibnamefont{and}
  \bibinfo{author}{\bibfnamefont{J.}~\bibnamefont{Goold}},
  \bibinfo{journal}{New Journal of Physics} \textbf{\bibinfo{volume}{17}},
  \bibinfo{pages}{075015} (\bibinfo{year}{2015}), ISSN
  \bibinfo{issn}{1367-2630}.

\bibitem[{\citenamefont{Ferraro et~al.}(2018)\citenamefont{Ferraro, Campisi,
  Andolina, Pellegrini, and Polini}}]{ferraro-2018}
\bibinfo{author}{\bibfnamefont{D.}~\bibnamefont{Ferraro}},
  \bibinfo{author}{\bibfnamefont{M.}~\bibnamefont{Campisi}},
  \bibinfo{author}{\bibfnamefont{G.~M.} \bibnamefont{Andolina}},
  \bibinfo{author}{\bibfnamefont{V.}~\bibnamefont{Pellegrini}},
  \bibnamefont{and} \bibinfo{author}{\bibfnamefont{M.}~\bibnamefont{Polini}},
  \bibinfo{journal}{Phys. Rev. Lett.} \textbf{\bibinfo{volume}{120}},
  \bibinfo{pages}{117702} (\bibinfo{year}{2018}).

\bibitem[{\citenamefont{Andolina et~al.}(2018)\citenamefont{Andolina, Farina,
  Mari, Pellegrini, Giovannetti, and Polini}}]{andolina-2018}
\bibinfo{author}{\bibfnamefont{G.~M.} \bibnamefont{Andolina}},
  \bibinfo{author}{\bibfnamefont{D.}~\bibnamefont{Farina}},
  \bibinfo{author}{\bibfnamefont{A.}~\bibnamefont{Mari}},
  \bibinfo{author}{\bibfnamefont{V.}~\bibnamefont{Pellegrini}},
  \bibinfo{author}{\bibfnamefont{V.}~\bibnamefont{Giovannetti}},
  \bibnamefont{and} \bibinfo{author}{\bibfnamefont{M.}~\bibnamefont{Polini}},
  \bibinfo{journal}{Phys. Rev. B} \textbf{\bibinfo{volume}{98}},
  \bibinfo{pages}{205423} (\bibinfo{year}{2018}),
  \urlprefix\url{https://link.aps.org/doi/10.1103/PhysRevB.98.205423}.

\bibitem[{\citenamefont{Liu et~al.}(2019)\citenamefont{Liu, Segal, and
  Hanna}}]{liu-2019}
\bibinfo{author}{\bibfnamefont{J.}~\bibnamefont{Liu}},
  \bibinfo{author}{\bibfnamefont{D.}~\bibnamefont{Segal}}, \bibnamefont{and}
  \bibinfo{author}{\bibfnamefont{G.}~\bibnamefont{Hanna}},
  \bibinfo{journal}{The Journal of Physical Chemistry C}
  \textbf{\bibinfo{volume}{123}}, \bibinfo{pages}{18303}
  (\bibinfo{year}{2019}).

\bibitem[{\citenamefont{Rossini et~al.}(2019)\citenamefont{Rossini, Andolina,
  and Polini}}]{rossini-2019}
\bibinfo{author}{\bibfnamefont{D.}~\bibnamefont{Rossini}},
  \bibinfo{author}{\bibfnamefont{G.~M.} \bibnamefont{Andolina}},
  \bibnamefont{and} \bibinfo{author}{\bibfnamefont{M.}~\bibnamefont{Polini}},
  \bibinfo{journal}{Phys. Rev. B} \textbf{\bibinfo{volume}{100}},
  \bibinfo{pages}{115142} (\bibinfo{year}{2019}).

\bibitem[{\citenamefont{Quach et~al.}(2022)\citenamefont{Quach, McGhee, Ganzer,
  Rouse, Lovett, Gauger, Keeling, Cerullo, Lidzey, and Virgili}}]{quach-2022}
\bibinfo{author}{\bibfnamefont{J.~Q.} \bibnamefont{Quach}},
  \bibinfo{author}{\bibfnamefont{K.~E.} \bibnamefont{McGhee}},
  \bibinfo{author}{\bibfnamefont{L.}~\bibnamefont{Ganzer}},
  \bibinfo{author}{\bibfnamefont{D.~M.} \bibnamefont{Rouse}},
  \bibinfo{author}{\bibfnamefont{B.~W.} \bibnamefont{Lovett}},
  \bibinfo{author}{\bibfnamefont{E.~M.} \bibnamefont{Gauger}},
  \bibinfo{author}{\bibfnamefont{J.}~\bibnamefont{Keeling}},
  \bibinfo{author}{\bibfnamefont{G.}~\bibnamefont{Cerullo}},
  \bibinfo{author}{\bibfnamefont{D.~G.} \bibnamefont{Lidzey}},
  \bibnamefont{and} \bibinfo{author}{\bibfnamefont{T.}~\bibnamefont{Virgili}},
  \bibinfo{journal}{Science Advances} \textbf{\bibinfo{volume}{8}},
  \bibinfo{pages}{eabk3160} (\bibinfo{year}{2022}).

\bibitem[{\citenamefont{Ahmadi et~al.}(2024)\citenamefont{Ahmadi, Mazurek,
  Horodecki, and Barzanjeh}}]{barzanjeh-2024}
\bibinfo{author}{\bibfnamefont{B.}~\bibnamefont{Ahmadi}},
  \bibinfo{author}{\bibfnamefont{P.}~\bibnamefont{Mazurek}},
  \bibinfo{author}{\bibfnamefont{P.}~\bibnamefont{Horodecki}},
  \bibnamefont{and}
  \bibinfo{author}{\bibfnamefont{S.}~\bibnamefont{Barzanjeh}},
  \bibinfo{journal}{Phys. Rev. Lett.} \textbf{\bibinfo{volume}{132}},
  \bibinfo{pages}{210402} (\bibinfo{year}{2024}),
  \urlprefix\url{https://link.aps.org/doi/10.1103/PhysRevLett.132.210402}.

\bibitem[{\citenamefont{Campaioli et~al.}(2024)\citenamefont{Campaioli,
  Gherardini, Quach, Polini, and Andolina}}]{campaioli-2024}
\bibinfo{author}{\bibfnamefont{F.}~\bibnamefont{Campaioli}},
  \bibinfo{author}{\bibfnamefont{S.}~\bibnamefont{Gherardini}},
  \bibinfo{author}{\bibfnamefont{J.~Q.} \bibnamefont{Quach}},
  \bibinfo{author}{\bibfnamefont{M.}~\bibnamefont{Polini}}, \bibnamefont{and}
  \bibinfo{author}{\bibfnamefont{G.~M.} \bibnamefont{Andolina}},
  \bibinfo{journal}{Rev. Mod. Phys.} \textbf{\bibinfo{volume}{96}},
  \bibinfo{pages}{031001} (\bibinfo{year}{2024}).

\bibitem[{\citenamefont{Hovhannisyan et~al.}(2013)\citenamefont{Hovhannisyan,
  Perarnau-Llobet, Huber, and Ac\'{\i}n}}]{hovhannisyan-2013}
\bibinfo{author}{\bibfnamefont{K.~V.} \bibnamefont{Hovhannisyan}},
  \bibinfo{author}{\bibfnamefont{M.}~\bibnamefont{Perarnau-Llobet}},
  \bibinfo{author}{\bibfnamefont{M.}~\bibnamefont{Huber}}, \bibnamefont{and}
  \bibinfo{author}{\bibfnamefont{A.}~\bibnamefont{Ac\'{\i}n}},
  \bibinfo{journal}{Phys. Rev. Lett.} \textbf{\bibinfo{volume}{111}},
  \bibinfo{pages}{240401} (\bibinfo{year}{2013}).

\bibitem[{\citenamefont{Caravelli et~al.}(2020)\citenamefont{Caravelli,
  Coulter-De~Wit, Garc\'{\i}a-Pintos, and Hamma}}]{caravelli-2020}
\bibinfo{author}{\bibfnamefont{F.}~\bibnamefont{Caravelli}},
  \bibinfo{author}{\bibfnamefont{G.}~\bibnamefont{Coulter-De~Wit}},
  \bibinfo{author}{\bibfnamefont{L.~P.} \bibnamefont{Garc\'{\i}a-Pintos}},
  \bibnamefont{and} \bibinfo{author}{\bibfnamefont{A.}~\bibnamefont{Hamma}},
  \bibinfo{journal}{Phys. Rev. Res.} \textbf{\bibinfo{volume}{2}},
  \bibinfo{pages}{023095} (\bibinfo{year}{2020}),
  \urlprefix\url{https://link.aps.org/doi/10.1103/PhysRevResearch.2.023095}.

\bibitem[{\citenamefont{Sachdev and Ye}(1993)}]{sachdev-1993}
\bibinfo{author}{\bibfnamefont{S.}~\bibnamefont{Sachdev}} \bibnamefont{and}
  \bibinfo{author}{\bibfnamefont{J.}~\bibnamefont{Ye}}, \bibinfo{journal}{Phys.
  Rev. Lett.} \textbf{\bibinfo{volume}{70}}, \bibinfo{pages}{3339}
  (\bibinfo{year}{1993}).

\bibitem[{\citenamefont{Kitaev}(2015{\natexlab{a}})}]{kitaev-2015-p1}
\bibinfo{author}{\bibfnamefont{A.}~\bibnamefont{Kitaev}},
  \emph{\bibinfo{title}{A simple model of quantum holography (part 1)}},
  \bibinfo{howpublished}{\url{https://online.kitp.ucsb.edu/online/entangled15/kitaev/}}
  (\bibinfo{year}{2015}{\natexlab{a}}), \bibinfo{note}{talk at KITP, University
  of California, Santa Barbara]}.

\bibitem[{\citenamefont{Kitaev}(2015{\natexlab{b}})}]{kitaev-2015-p2}
\bibinfo{author}{\bibfnamefont{A.}~\bibnamefont{Kitaev}},
  \emph{\bibinfo{title}{A simple model of quantum holography (part 2)}},
  \bibinfo{howpublished}{\url{https://online.kitp.ucsb.edu/online/entangled15/kitaev2/}}
  (\bibinfo{year}{2015}{\natexlab{b}}), \bibinfo{note}{talk at KITP, University
  of California, Santa Barbara}.

\bibitem[{\citenamefont{Sachdev}(2015)}]{sachdev-2015}
\bibinfo{author}{\bibfnamefont{S.}~\bibnamefont{Sachdev}},
  \bibinfo{journal}{Phys. Rev. X} \textbf{\bibinfo{volume}{5}},
  \bibinfo{pages}{041025} (\bibinfo{year}{2015}).

\bibitem[{\citenamefont{Gu et~al.}(2020)\citenamefont{Gu, Kitaev, Sachdev, and
  Tarnopolsky}}]{gu-2020}
\bibinfo{author}{\bibfnamefont{Y.}~\bibnamefont{Gu}},
  \bibinfo{author}{\bibfnamefont{A.}~\bibnamefont{Kitaev}},
  \bibinfo{author}{\bibfnamefont{S.}~\bibnamefont{Sachdev}}, \bibnamefont{and}
  \bibinfo{author}{\bibfnamefont{G.}~\bibnamefont{Tarnopolsky}},
  \bibinfo{journal}{Journal of High Energy Physics}
  \textbf{\bibinfo{volume}{2020}} (\bibinfo{year}{2020}).

\bibitem[{\citenamefont{Belyansky et~al.}(2020)\citenamefont{Belyansky,
  Bienias, Kharkov, Gorshkov, and Swingle}}]{swingle-2020}
\bibinfo{author}{\bibfnamefont{R.}~\bibnamefont{Belyansky}},
  \bibinfo{author}{\bibfnamefont{P.}~\bibnamefont{Bienias}},
  \bibinfo{author}{\bibfnamefont{Y.~A.} \bibnamefont{Kharkov}},
  \bibinfo{author}{\bibfnamefont{A.~V.} \bibnamefont{Gorshkov}},
  \bibnamefont{and} \bibinfo{author}{\bibfnamefont{B.}~\bibnamefont{Swingle}},
  \bibinfo{journal}{Physical Review Letters} \textbf{\bibinfo{volume}{125}}
  (\bibinfo{year}{2020}), ISSN \bibinfo{issn}{1079-7114}.

\bibitem[{\citenamefont{Rossini et~al.}(2020)\citenamefont{Rossini, Andolina,
  Rosa, Carrega, and Polini}}]{rossini-2020}
\bibinfo{author}{\bibfnamefont{D.}~\bibnamefont{Rossini}},
  \bibinfo{author}{\bibfnamefont{G.~M.} \bibnamefont{Andolina}},
  \bibinfo{author}{\bibfnamefont{D.}~\bibnamefont{Rosa}},
  \bibinfo{author}{\bibfnamefont{M.}~\bibnamefont{Carrega}}, \bibnamefont{and}
  \bibinfo{author}{\bibfnamefont{M.}~\bibnamefont{Polini}},
  \bibinfo{journal}{Phys. Rev. Lett.} \textbf{\bibinfo{volume}{125}},
  \bibinfo{pages}{236402} (\bibinfo{year}{2020}).

\bibitem[{\citenamefont{Rosa et~al.}(2020)\citenamefont{Rosa, Rossini,
  Andolina, Polini, and Carrega}}]{rosa-2020}
\bibinfo{author}{\bibfnamefont{D.}~\bibnamefont{Rosa}},
  \bibinfo{author}{\bibfnamefont{D.}~\bibnamefont{Rossini}},
  \bibinfo{author}{\bibfnamefont{G.~M.} \bibnamefont{Andolina}},
  \bibinfo{author}{\bibfnamefont{M.}~\bibnamefont{Polini}}, \bibnamefont{and}
  \bibinfo{author}{\bibfnamefont{M.}~\bibnamefont{Carrega}},
  \bibinfo{journal}{Journal of High Energy Physics}
  \textbf{\bibinfo{volume}{2020}} (\bibinfo{year}{2020}), ISSN
  \bibinfo{issn}{1029-8479}.

\bibitem[{\citenamefont{Danshita et~al.}(2017)\citenamefont{Danshita, Hanada,
  and Tezuka}}]{danshita-2017}
\bibinfo{author}{\bibfnamefont{I.}~\bibnamefont{Danshita}},
  \bibinfo{author}{\bibfnamefont{M.}~\bibnamefont{Hanada}}, \bibnamefont{and}
  \bibinfo{author}{\bibfnamefont{M.}~\bibnamefont{Tezuka}},
  \bibinfo{journal}{Progress of Theoretical and Experimental Physics}
  \textbf{\bibinfo{volume}{2017}} (\bibinfo{year}{2017}), ISSN
  \bibinfo{issn}{2050-3911}.

\bibitem[{\citenamefont{Chen et~al.}(2018)\citenamefont{Chen, Ilan, de~Juan,
  Pikulin, and Franz}}]{chen-2018}
\bibinfo{author}{\bibfnamefont{A.}~\bibnamefont{Chen}},
  \bibinfo{author}{\bibfnamefont{R.}~\bibnamefont{Ilan}},
  \bibinfo{author}{\bibfnamefont{F.}~\bibnamefont{de~Juan}},
  \bibinfo{author}{\bibfnamefont{D.~I.} \bibnamefont{Pikulin}},
  \bibnamefont{and} \bibinfo{author}{\bibfnamefont{M.}~\bibnamefont{Franz}},
  \bibinfo{journal}{Phys. Rev. Lett.} \textbf{\bibinfo{volume}{121}},
  \bibinfo{pages}{036403} (\bibinfo{year}{2018}).

\bibitem[{\citenamefont{Can et~al.}(2019)\citenamefont{Can, Nica, and
  Franz}}]{can-2019}
\bibinfo{author}{\bibfnamefont{O.}~\bibnamefont{Can}},
  \bibinfo{author}{\bibfnamefont{E.~M.} \bibnamefont{Nica}}, \bibnamefont{and}
  \bibinfo{author}{\bibfnamefont{M.}~\bibnamefont{Franz}},
  \bibinfo{journal}{Phys. Rev. B} \textbf{\bibinfo{volume}{99}},
  \bibinfo{pages}{045419} (\bibinfo{year}{2019}).

\bibitem[{\citenamefont{Uhrich et~al.}(2023)\citenamefont{Uhrich,
  Bandyopadhyay, Sauerwein, Sonner, Brantut, and Hauke}}]{hauke-2023}
\bibinfo{author}{\bibfnamefont{P.}~\bibnamefont{Uhrich}},
  \bibinfo{author}{\bibfnamefont{S.}~\bibnamefont{Bandyopadhyay}},
  \bibinfo{author}{\bibfnamefont{N.}~\bibnamefont{Sauerwein}},
  \bibinfo{author}{\bibfnamefont{J.}~\bibnamefont{Sonner}},
  \bibinfo{author}{\bibfnamefont{J.-P.} \bibnamefont{Brantut}},
  \bibnamefont{and} \bibinfo{author}{\bibfnamefont{P.}~\bibnamefont{Hauke}},
  \emph{\bibinfo{title}{A cavity quantum electrodynamics implementation of the
  sachdev--ye--kitaev model}} (\bibinfo{year}{2023}), \eprint{2303.11343}.

\bibitem[{\citenamefont{Baumgartner et~al.}(2024)\citenamefont{Baumgartner,
  Pelliconi, Bandyopadhyay, Orsi, Sauerwein, Hauke, Brantut, and
  Sonner}}]{hauke-2024}
\bibinfo{author}{\bibfnamefont{R.}~\bibnamefont{Baumgartner}},
  \bibinfo{author}{\bibfnamefont{P.}~\bibnamefont{Pelliconi}},
  \bibinfo{author}{\bibfnamefont{S.}~\bibnamefont{Bandyopadhyay}},
  \bibinfo{author}{\bibfnamefont{F.}~\bibnamefont{Orsi}},
  \bibinfo{author}{\bibfnamefont{N.}~\bibnamefont{Sauerwein}},
  \bibinfo{author}{\bibfnamefont{P.}~\bibnamefont{Hauke}},
  \bibinfo{author}{\bibfnamefont{J.-P.} \bibnamefont{Brantut}},
  \bibnamefont{and} \bibinfo{author}{\bibfnamefont{J.}~\bibnamefont{Sonner}},
  \emph{\bibinfo{title}{Quantum simulation of the sachdev-ye-kitaev model using
  time-dependent disorder in optical cavities}} (\bibinfo{year}{2024}),
  \eprint{2411.17802}.

\bibitem[{\citenamefont{Bettaque and Swingle}(2024)}]{bettaque-2024}
\bibinfo{author}{\bibfnamefont{V.}~\bibnamefont{Bettaque}} \bibnamefont{and}
  \bibinfo{author}{\bibfnamefont{B.}~\bibnamefont{Swingle}},
  \bibinfo{journal}{Quantum} \textbf{\bibinfo{volume}{8}},
  \bibinfo{pages}{1362} (\bibinfo{year}{2024}), ISSN \bibinfo{issn}{2521-327X},
  \urlprefix\url{http://dx.doi.org/10.22331/q-2024-05-27-1362}.

\bibitem[{\citenamefont{Fu and Sachdev}(2016)}]{fu-2016}
\bibinfo{author}{\bibfnamefont{W.}~\bibnamefont{Fu}} \bibnamefont{and}
  \bibinfo{author}{\bibfnamefont{S.}~\bibnamefont{Sachdev}},
  \bibinfo{journal}{Phys. Rev. B} \textbf{\bibinfo{volume}{94}},
  \bibinfo{pages}{035135} (\bibinfo{year}{2016}),
  \urlprefix\url{https://link.aps.org/doi/10.1103/PhysRevB.94.035135}.

\bibitem[{\citenamefont{Liu et~al.}(2018)\citenamefont{Liu, Chen, and
  Balents}}]{balents-2018}
\bibinfo{author}{\bibfnamefont{C.}~\bibnamefont{Liu}},
  \bibinfo{author}{\bibfnamefont{X.}~\bibnamefont{Chen}}, \bibnamefont{and}
  \bibinfo{author}{\bibfnamefont{L.}~\bibnamefont{Balents}},
  \bibinfo{journal}{Phys. Rev. B} \textbf{\bibinfo{volume}{97}},
  \bibinfo{pages}{245126} (\bibinfo{year}{2018}).

\bibitem[{\citenamefont{Zhang}(2022)}]{zhang-2022}
\bibinfo{author}{\bibfnamefont{P.}~\bibnamefont{Zhang}},
  \bibinfo{journal}{Frontiers of Physics} \textbf{\bibinfo{volume}{17}},
  \bibinfo{pages}{43201} (\bibinfo{year}{2022}).

\bibitem[{\citenamefont{Xu et~al.}(2020)\citenamefont{Xu, Susskind, Su, and
  Swingle}}]{xu-2020}
\bibinfo{author}{\bibfnamefont{S.}~\bibnamefont{Xu}},
  \bibinfo{author}{\bibfnamefont{L.}~\bibnamefont{Susskind}},
  \bibinfo{author}{\bibfnamefont{Y.}~\bibnamefont{Su}}, \bibnamefont{and}
  \bibinfo{author}{\bibfnamefont{B.}~\bibnamefont{Swingle}},
  \emph{\bibinfo{title}{A sparse model of quantum holography}}
  (\bibinfo{year}{2020}), \eprint{2008.02303}.

\bibitem[{\citenamefont{García-García
  et~al.}(2021)\citenamefont{García-García, Jia, Rosa, and
  Verbaarschot}}]{garcia-garcia-2021}
\bibinfo{author}{\bibfnamefont{A.~M.} \bibnamefont{García-García}},
  \bibinfo{author}{\bibfnamefont{Y.}~\bibnamefont{Jia}},
  \bibinfo{author}{\bibfnamefont{D.}~\bibnamefont{Rosa}}, \bibnamefont{and}
  \bibinfo{author}{\bibfnamefont{J.~J.} \bibnamefont{Verbaarschot}},
  \bibinfo{journal}{Physical Review D} \textbf{\bibinfo{volume}{103}}
  (\bibinfo{year}{2021}), ISSN \bibinfo{issn}{2470-0029}.

\bibitem[{\citenamefont{Orman et~al.}(2025)\citenamefont{Orman, Gharibyan, and
  Preskill}}]{preskill-2024}
\bibinfo{author}{\bibfnamefont{P.}~\bibnamefont{Orman}},
  \bibinfo{author}{\bibfnamefont{H.}~\bibnamefont{Gharibyan}},
  \bibnamefont{and} \bibinfo{author}{\bibfnamefont{J.}~\bibnamefont{Preskill}},
  \bibinfo{journal}{Journal of High Energy Physics}
  \textbf{\bibinfo{volume}{2025}} (\bibinfo{year}{2025}).

\bibitem[{\citenamefont{Sisorio et~al.}(2026)\citenamefont{Sisorio, Cappellaro,
  and Dell'Anna}}]{SM}
\bibinfo{author}{\bibfnamefont{G.}~\bibnamefont{Sisorio}},
  \bibinfo{author}{\bibfnamefont{A.}~\bibnamefont{Cappellaro}},
  \bibnamefont{and}
  \bibinfo{author}{\bibfnamefont{L.}~\bibnamefont{Dell'Anna}},
  \emph{\bibinfo{title}{Supplemental material: Sparse syk quantum batteries:
  Boosting quantum efficiency by reducing complexity}} (\bibinfo{year}{2026}).

\bibitem[{\citenamefont{Maldacena et~al.}(2016)\citenamefont{Maldacena,
  Shenker, and Stanford}}]{maldacena-2016}
\bibinfo{author}{\bibfnamefont{J.}~\bibnamefont{Maldacena}},
  \bibinfo{author}{\bibfnamefont{S.~H.} \bibnamefont{Shenker}},
  \bibnamefont{and} \bibinfo{author}{\bibfnamefont{D.}~\bibnamefont{Stanford}},
  \bibinfo{journal}{Journal of High Energy Physics}
  \textbf{\bibinfo{volume}{2016}} (\bibinfo{year}{2016}), ISSN
  \bibinfo{issn}{1029-8479}.

\bibitem[{\citenamefont{Guhr et~al.}(1998)\citenamefont{Guhr,
  Müller–Groeling, and Weidenmüller}}]{guhr-1998}
\bibinfo{author}{\bibfnamefont{T.}~\bibnamefont{Guhr}},
  \bibinfo{author}{\bibfnamefont{A.}~\bibnamefont{Müller–Groeling}},
  \bibnamefont{and} \bibinfo{author}{\bibfnamefont{H.~A.}
  \bibnamefont{Weidenmüller}}, \bibinfo{journal}{Physics Reports}
  \textbf{\bibinfo{volume}{299}}, \bibinfo{pages}{189–425}
  (\bibinfo{year}{1998}), ISSN \bibinfo{issn}{0370-1573}.

\bibitem[{\citenamefont{D’Alessio et~al.}(2016)\citenamefont{D’Alessio,
  Kafri, Polkovnikov, and Rigol}}]{dalessio-2016}
\bibinfo{author}{\bibfnamefont{L.}~\bibnamefont{D’Alessio}},
  \bibinfo{author}{\bibfnamefont{Y.}~\bibnamefont{Kafri}},
  \bibinfo{author}{\bibfnamefont{A.}~\bibnamefont{Polkovnikov}},
  \bibnamefont{and} \bibinfo{author}{\bibfnamefont{M.}~\bibnamefont{Rigol}},
  \bibinfo{journal}{Advances in Physics} \textbf{\bibinfo{volume}{65}},
  \bibinfo{pages}{239–362} (\bibinfo{year}{2016}), ISSN
  \bibinfo{issn}{1460-6976}.

\bibitem[{\citenamefont{You et~al.}(2017)\citenamefont{You, Ludwig, and
  Xu}}]{ludwig-2017}
\bibinfo{author}{\bibfnamefont{Y.-Z.} \bibnamefont{You}},
  \bibinfo{author}{\bibfnamefont{A.~W.~W.} \bibnamefont{Ludwig}},
  \bibnamefont{and} \bibinfo{author}{\bibfnamefont{C.}~\bibnamefont{Xu}},
  \bibinfo{journal}{Phys. Rev. B} \textbf{\bibinfo{volume}{95}},
  \bibinfo{pages}{115150} (\bibinfo{year}{2017}),
  \urlprefix\url{https://link.aps.org/doi/10.1103/PhysRevB.95.115150}.

\bibitem[{\citenamefont{Allahverdyan et~al.}(2004)\citenamefont{Allahverdyan,
  Balian, and Nieuwenhuizen}}]{allahverdyan-2004}
\bibinfo{author}{\bibfnamefont{A.~E.} \bibnamefont{Allahverdyan}},
  \bibinfo{author}{\bibfnamefont{R.}~\bibnamefont{Balian}}, \bibnamefont{and}
  \bibinfo{author}{\bibfnamefont{T.~M.} \bibnamefont{Nieuwenhuizen}},
  \bibinfo{journal}{Europhysics Letters} \textbf{\bibinfo{volume}{67}},
  \bibinfo{pages}{565} (\bibinfo{year}{2004}).

\bibitem[{\citenamefont{Tirone et~al.}(2021)\citenamefont{Tirone, Salvia, and
  Giovannetti}}]{tirone-2021}
\bibinfo{author}{\bibfnamefont{S.}~\bibnamefont{Tirone}},
  \bibinfo{author}{\bibfnamefont{R.}~\bibnamefont{Salvia}}, \bibnamefont{and}
  \bibinfo{author}{\bibfnamefont{V.}~\bibnamefont{Giovannetti}},
  \bibinfo{journal}{Physical Review Letters} \textbf{\bibinfo{volume}{127}}
  (\bibinfo{year}{2021}), ISSN \bibinfo{issn}{1079-7114}.

\bibitem[{\citenamefont{Touil et~al.}(2021)\citenamefont{Touil, Çakmak, and
  Deffner}}]{touil-2021}
\bibinfo{author}{\bibfnamefont{A.}~\bibnamefont{Touil}},
  \bibinfo{author}{\bibfnamefont{B.}~\bibnamefont{Çakmak}}, \bibnamefont{and}
  \bibinfo{author}{\bibfnamefont{S.}~\bibnamefont{Deffner}},
  \bibinfo{journal}{Journal of Physics A: Mathematical and Theoretical}
  \textbf{\bibinfo{volume}{55}}, \bibinfo{pages}{025301}
  (\bibinfo{year}{2021}), ISSN \bibinfo{issn}{1751-8121}.

\bibitem[{\citenamefont{Lenard}(1978)}]{lenard-1978}
\bibinfo{author}{\bibfnamefont{A.}~\bibnamefont{Lenard}}, \bibinfo{journal}{J.
  Statist. Phys.} \textbf{\bibinfo{volume}{19}}, \bibinfo{pages}{575}
  (\bibinfo{year}{1978}).

\bibitem[{\citenamefont{Julià-Farré et~al.}(2020)\citenamefont{Julià-Farré,
  Salamon, Riera, Bera, and Lewenstein}}]{julia-2020}
\bibinfo{author}{\bibfnamefont{S.}~\bibnamefont{Julià-Farré}},
  \bibinfo{author}{\bibfnamefont{T.}~\bibnamefont{Salamon}},
  \bibinfo{author}{\bibfnamefont{R.}~\bibnamefont{Riera}},
  \bibinfo{author}{\bibfnamefont{M.~N.} \bibnamefont{Bera}}, \bibnamefont{and}
  \bibinfo{author}{\bibfnamefont{M.}~\bibnamefont{Lewenstein}},
  \bibinfo{journal}{Phys. Rev. Research} \textbf{\bibinfo{volume}{2}},
  \bibinfo{pages}{023113} (\bibinfo{year}{2020}).

\bibitem[{\citenamefont{Brzezi\ifmmode~\acute{n}\else \'{n}\fi{}ska
  et~al.}(2023)\citenamefont{Brzezi\ifmmode~\acute{n}\else \'{n}\fi{}ska, Guan,
  Yazyev, Sachdev, and Kruchkov}}]{kruchkov-2023}
\bibinfo{author}{\bibfnamefont{M.}~\bibnamefont{Brzezi\ifmmode~\acute{n}\else
  \'{n}\fi{}ska}}, \bibinfo{author}{\bibfnamefont{Y.}~\bibnamefont{Guan}},
  \bibinfo{author}{\bibfnamefont{O.~V.} \bibnamefont{Yazyev}},
  \bibinfo{author}{\bibfnamefont{S.}~\bibnamefont{Sachdev}}, \bibnamefont{and}
  \bibinfo{author}{\bibfnamefont{A.}~\bibnamefont{Kruchkov}},
  \bibinfo{journal}{Phys. Rev. Lett.} \textbf{\bibinfo{volume}{131}},
  \bibinfo{pages}{036503} (\bibinfo{year}{2023}).

\bibitem[{\citenamefont{Sisorio et~al.}(2025)\citenamefont{Sisorio, Cappellaro,
  and Dell'Anna}}]{sisorio-2025}
\bibinfo{author}{\bibfnamefont{G.}~\bibnamefont{Sisorio}},
  \bibinfo{author}{\bibfnamefont{A.}~\bibnamefont{Cappellaro}},
  \bibnamefont{and}
  \bibinfo{author}{\bibfnamefont{L.}~\bibnamefont{Dell'Anna}},
  \emph{\bibinfo{title}{Sparse syk quantum batteries: Boosting quantum
  efficiency by reducing complexity}} (\bibinfo{year}{2025}),
  \urlprefix\url{https://doi.org/10.5281/zenodo.15561904}.

\bibitem[{\citenamefont{Johansson et~al.}(2012)\citenamefont{Johansson, Nation,
  and Nori}}]{johansson-2012}
\bibinfo{author}{\bibfnamefont{J.}~\bibnamefont{Johansson}},
  \bibinfo{author}{\bibfnamefont{P.}~\bibnamefont{Nation}}, \bibnamefont{and}
  \bibinfo{author}{\bibfnamefont{F.}~\bibnamefont{Nori}},
  \bibinfo{journal}{Computer Physics Communications}
  \textbf{\bibinfo{volume}{183}}, \bibinfo{pages}{1760} (\bibinfo{year}{2012}),
  ISSN \bibinfo{issn}{0010-4655},
  \urlprefix\url{https://www.sciencedirect.com/science/article/pii/S0010465512000835}.

\bibitem[{\citenamefont{Johansson et~al.}(2013)\citenamefont{Johansson, Nation,
  and Nori}}]{johansson-2013}
\bibinfo{author}{\bibfnamefont{J.}~\bibnamefont{Johansson}},
  \bibinfo{author}{\bibfnamefont{P.}~\bibnamefont{Nation}}, \bibnamefont{and}
  \bibinfo{author}{\bibfnamefont{F.}~\bibnamefont{Nori}},
  \bibinfo{journal}{Computer Physics Communications}
  \textbf{\bibinfo{volume}{184}}, \bibinfo{pages}{1234} (\bibinfo{year}{2013}),
  ISSN \bibinfo{issn}{0010-4655},
  \urlprefix\url{https://www.sciencedirect.com/science/article/pii/S0010465512003955}.

\bibitem[{\citenamefont{Lambert et~al.}(2024)\citenamefont{Lambert, Giguère,
  Menczel, Li, Hopf, Suárez, Gali, Lishman, Gadhvi, Agarwal
  et~al.}}]{lambert-2024-qutip5}
\bibinfo{author}{\bibfnamefont{N.}~\bibnamefont{Lambert}},
  \bibinfo{author}{\bibfnamefont{E.}~\bibnamefont{Giguère}},
  \bibinfo{author}{\bibfnamefont{P.}~\bibnamefont{Menczel}},
  \bibinfo{author}{\bibfnamefont{B.}~\bibnamefont{Li}},
  \bibinfo{author}{\bibfnamefont{P.}~\bibnamefont{Hopf}},
  \bibinfo{author}{\bibfnamefont{G.}~\bibnamefont{Suárez}},
  \bibinfo{author}{\bibfnamefont{M.}~\bibnamefont{Gali}},
  \bibinfo{author}{\bibfnamefont{J.}~\bibnamefont{Lishman}},
  \bibinfo{author}{\bibfnamefont{R.}~\bibnamefont{Gadhvi}},
  \bibinfo{author}{\bibfnamefont{R.}~\bibnamefont{Agarwal}},
  \bibnamefont{et~al.}, \emph{\bibinfo{title}{Qutip 5: The quantum toolbox in
  python}} (\bibinfo{year}{2024}), \eprint{2412.04705},
  \urlprefix\url{https://arxiv.org/abs/2412.04705}.

\bibitem[{\citenamefont{Garc\'{\i}a-\'Alvarez
  et~al.}(2017)\citenamefont{Garc\'{\i}a-\'Alvarez, Egusquiza, Lamata, del
  Campo, Sonner, and Solano}}]{solano-2017}
\bibinfo{author}{\bibfnamefont{L.}~\bibnamefont{Garc\'{\i}a-\'Alvarez}},
  \bibinfo{author}{\bibfnamefont{I.~L.} \bibnamefont{Egusquiza}},
  \bibinfo{author}{\bibfnamefont{L.}~\bibnamefont{Lamata}},
  \bibinfo{author}{\bibfnamefont{A.}~\bibnamefont{del Campo}},
  \bibinfo{author}{\bibfnamefont{J.}~\bibnamefont{Sonner}}, \bibnamefont{and}
  \bibinfo{author}{\bibfnamefont{E.}~\bibnamefont{Solano}},
  \bibinfo{journal}{Phys. Rev. Lett.} \textbf{\bibinfo{volume}{119}},
  \bibinfo{pages}{040501} (\bibinfo{year}{2017}),
  \urlprefix\url{https://link.aps.org/doi/10.1103/PhysRevLett.119.040501}.

\bibitem[{\citenamefont{Sonner and Vielma}(2017)}]{sonner-2017}
\bibinfo{author}{\bibfnamefont{J.}~\bibnamefont{Sonner}} \bibnamefont{and}
  \bibinfo{author}{\bibfnamefont{M.}~\bibnamefont{Vielma}},
  \bibinfo{journal}{Journal of High Energy Physics}
  \textbf{\bibinfo{volume}{2017}} (\bibinfo{year}{2017}), ISSN
  \bibinfo{issn}{1029-8479}.

\bibitem[{\citenamefont{Saad et~al.}(2019)\citenamefont{Saad, Shenker, and
  Stanford}}]{saad-2019}
\bibinfo{author}{\bibfnamefont{P.}~\bibnamefont{Saad}},
  \bibinfo{author}{\bibfnamefont{S.~H.} \bibnamefont{Shenker}},
  \bibnamefont{and} \bibinfo{author}{\bibfnamefont{D.}~\bibnamefont{Stanford}},
  \emph{\bibinfo{title}{A semiclassical ramp in syk and in gravity}}
  (\bibinfo{year}{2019}), \eprint{1806.06840}.

\bibitem[{\citenamefont{Hartnoll et~al.}(2018)\citenamefont{Hartnoll, Lucas,
  and Sachdev}}]{hartnoll-2018}
\bibinfo{author}{\bibfnamefont{S.~A.} \bibnamefont{Hartnoll}},
  \bibinfo{author}{\bibfnamefont{A.}~\bibnamefont{Lucas}}, \bibnamefont{and}
  \bibinfo{author}{\bibfnamefont{S.}~\bibnamefont{Sachdev}},
  \emph{\bibinfo{title}{Holographic quantum matter}} (\bibinfo{year}{2018}),
  \eprint{1612.07324}, \urlprefix\url{https://arxiv.org/abs/1612.07324}.

\bibitem[{\citenamefont{Eberlein et~al.}(2017)\citenamefont{Eberlein, Kasper,
  Sachdev, and Steinberg}}]{eberlein-2017}
\bibinfo{author}{\bibfnamefont{A.}~\bibnamefont{Eberlein}},
  \bibinfo{author}{\bibfnamefont{V.}~\bibnamefont{Kasper}},
  \bibinfo{author}{\bibfnamefont{S.}~\bibnamefont{Sachdev}}, \bibnamefont{and}
  \bibinfo{author}{\bibfnamefont{J.}~\bibnamefont{Steinberg}},
  \bibinfo{journal}{Phys. Rev. B} \textbf{\bibinfo{volume}{96}},
  \bibinfo{pages}{205123} (\bibinfo{year}{2017}).

\bibitem[{\citenamefont{Dicke}(1954)}]{dicke-1954}
\bibinfo{author}{\bibfnamefont{R.~H.} \bibnamefont{Dicke}},
  \bibinfo{journal}{Phys. Rev.} \textbf{\bibinfo{volume}{93}},
  \bibinfo{pages}{99} (\bibinfo{year}{1954}).

\bibitem[{\citenamefont{Kirton et~al.}(2018)\citenamefont{Kirton, Roses,
  Keeling, and Dalla~Torre}}]{kirton-2018}
\bibinfo{author}{\bibfnamefont{P.}~\bibnamefont{Kirton}},
  \bibinfo{author}{\bibfnamefont{M.~M.} \bibnamefont{Roses}},
  \bibinfo{author}{\bibfnamefont{J.}~\bibnamefont{Keeling}}, \bibnamefont{and}
  \bibinfo{author}{\bibfnamefont{E.~G.} \bibnamefont{Dalla~Torre}},
  \bibinfo{journal}{Advanced Quantum Technologies} \textbf{\bibinfo{volume}{2}}
  (\bibinfo{year}{2018}), ISSN \bibinfo{issn}{2511-9044}.

\bibitem[{\citenamefont{Dukelsky et~al.}(2004)\citenamefont{Dukelsky, Dussel,
  Esebbag, and Pittel}}]{dukelsky-2004}
\bibinfo{author}{\bibfnamefont{J.}~\bibnamefont{Dukelsky}},
  \bibinfo{author}{\bibfnamefont{G.~G.} \bibnamefont{Dussel}},
  \bibinfo{author}{\bibfnamefont{C.}~\bibnamefont{Esebbag}}, \bibnamefont{and}
  \bibinfo{author}{\bibfnamefont{S.}~\bibnamefont{Pittel}},
  \bibinfo{journal}{Phys. Rev. Lett.} \textbf{\bibinfo{volume}{93}},
  \bibinfo{pages}{050403} (\bibinfo{year}{2004}).

\bibitem[{\citenamefont{Str\"ater et~al.}(2012)\citenamefont{Str\"ater,
  Tsyplyatyev, and Faribault}}]{strater-2012}
\bibinfo{author}{\bibfnamefont{C.}~\bibnamefont{Str\"ater}},
  \bibinfo{author}{\bibfnamefont{O.}~\bibnamefont{Tsyplyatyev}},
  \bibnamefont{and}
  \bibinfo{author}{\bibfnamefont{A.}~\bibnamefont{Faribault}},
  \bibinfo{journal}{Phys. Rev. B} \textbf{\bibinfo{volume}{86}},
  \bibinfo{pages}{195101} (\bibinfo{year}{2012}).

\end{thebibliography}

\section{Supplemental material}
\subsection{Jordan-Wigner transformation}
The complex SYK can be mapped onto a spin-$1/2$ model by means of a Jordan-Wigner
transformation, providing us with a more convenient numerical platform for our 
purposes \cite{solano-2017,rossini-2020}. By recalling 
$\hat{\sigma}^{\pm}_{i} = (\hat{\sigma}_i^x \pm i \hat{\sigma}_i^y)/2$, 
the transformation is defined by
\begin{equation}
\hat{c}^{\dagger}_i = \hat{\sigma}_i^{+} \prod_{m=1}^{i-1} \hat{\sigma}^z_m  \qquad \text{and}
\qquad
\hat{c}_i = \bigg(\prod_{m = 1}^{i-1} \sigma_m^z \bigg) \hat{\sigma}_i\;.
\label{jordan-wigner transformation}
\end{equation}
As noticed in \cite{rosa-2020}, this transformation make the highly nonlocal character
of the cSYK more transparent. As we aim to make clear later, this feature is particularly relevant when we compare the sparse cSYK to other battery implementations, such as, for instance, the Dicke-based one \cite{ferraro-2018}. 

\subsection{\textcolor{black}{Particle-hole symmetry}}
\textcolor{black}{
The Hamiltonian $\hat{H}_{\text{cSYK}}$, appearing in Eq.~(1) of the main text,
is invariant under particle-hole symmetry (PHS) in the
thermodynamic limit, namely for $N \rightarrow \infty$. 
For finite $N$,  $\hat{H}_{\text{cSYK}}$ should be supplemented by additional terms in order to enforce PHS \cite{fu-2016}, 
\begin{equation}
\hat{H}_{\text{cSYK}}\rightarrow \hat{H}_{\text{cSYK}}+\hat{H}_{\text{PHS}}
\end{equation}
with 
\begin{equation}
\hat{H}_{\text{PHS}}=\frac{1}{2}\sum_{i,j,k,l=1}^N J_{ijkl} (\delta_{ik}\hat c^\dagger_j \hat c_l+ \delta_{jl}\hat c^\dagger_i \hat c_k-\delta_{il}\hat c^\dagger_j \hat c_k+\delta_{jk}\hat c^\dagger_i \hat c_l)
\end{equation}
Since $\hat{H}_{\text{cSYK}}$ represents a charging protocol, rather than a real physical system, one can choose to work either with or without imposing PHS. 
However, we have checked that our main findings do not qualitatively change if PHS is not enforced and simply $\sum_{i,j,k,l = 1}^N J_{ijkl} \;\hat{c}^{\dagger}_i \hat{c}^{\dagger}_j\hat{c}_k\hat{c}_l $ is used as charging Hamiltonian, as also declared in \cite{rossini-2020} for the fully connected SYK batteries. 
}

\subsection{\textcolor{black}{Gap ratios and critical sparsity parameter}}
\textcolor{black}{
The gap ratio considered in Fig.~2 of the main text is obtained considering the spectrum of the Hamitonian $\hat{H}_{\text{cSYK}}$ in Eq.~(1) of the main text (its sparse version, with $J_{ijkl}\rightarrow x_{ijkl}J_{ijkl}$) without imposing PHS and with no constrain for the charge sector. The complex SYK model, actually commute with the number operator, therefore one can fix the occupation number, namely the number of fermions, $N_f$, sitting in $N$ sites. The spectrum, therefore, can be decomposed in different charge sectors. 
Actually fixing a charge sector, namely fixing $N_f$, the RMT dictates a Wigner-Dyson distribution of the energy spacing. 
In particular, for the complex SYK model the level statistics is generally that of a Gaussian unitary ensemble (GUE) \cite{ludwig-2017}, for which the gap ratio is $r\approx 0.6 $. If we consider all the charge sectors, the number of levels is much higher,  $\sum_{N_f=0}^N\binom{N}{N_f}=2^N$, and the energy levels are all mixed, therefore the average level spacing is expected to be smaller. 
\begin{figure}[h!]
    \centering
    \includegraphics[width=0.99\linewidth]{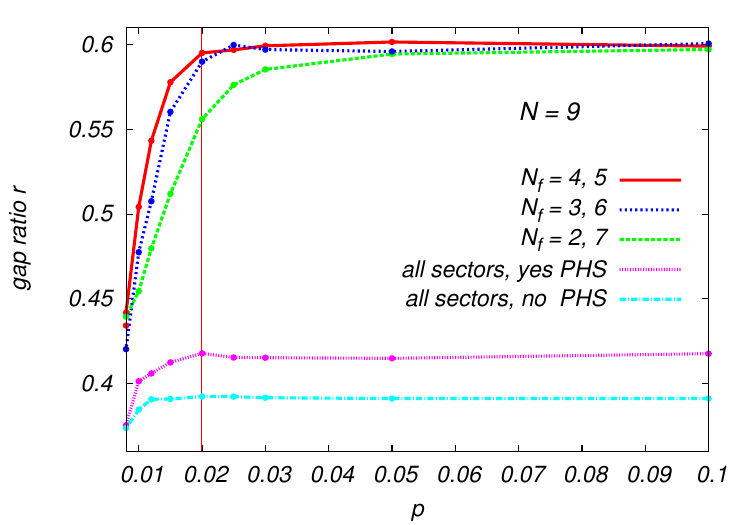}
    \caption{\textcolor{black}{Gap ratios for $N=9$ as functions of the sparsity parameter $p$, for different charge sectors, for the whole spectrum with mixed sectors preserving PHS and for the whole spectrum without PHS. The vertical line displays the $p_c$ value.}}
    \label{fig:r_N9}
\end{figure}
\begin{figure}[h!]
    \centering
    \includegraphics[width=0.99\linewidth]{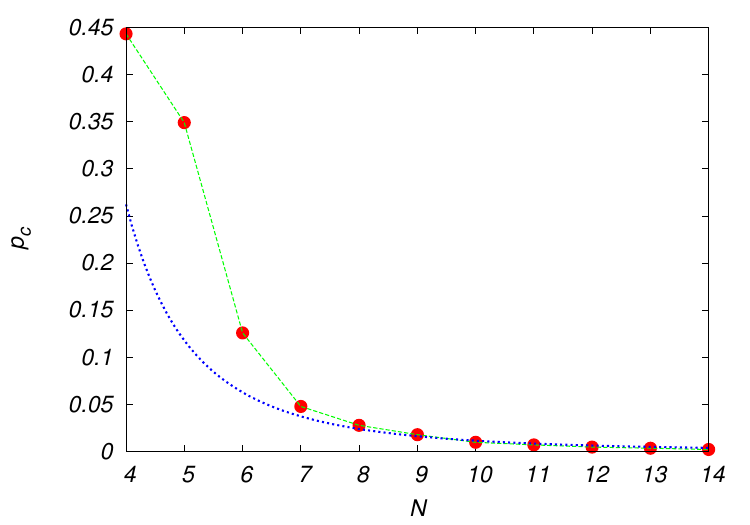}
    \caption{\textcolor{black}{The critical sparsity parameter $p_c$ for different values of $N$ (red dots, connected by the dashed green line). The dotted blue line is the curve $p_c=\frac{k_cN}{\binom{N}{2}^2}$, with $k_c\approx 2.4$.}}
    \label{fig:pc}
\end{figure}
\\
In Fig. \ref{fig:r_N9} we show, indeed that the gap ration, for $N=9$, is $r\approx 0.6 $ for any large sparsity parameter $p$ up to $p=1$ (not shown), for different charge sectors and imposing PHS (in fact the plots for $N_f$ and $(N-N_f)$ are the same). 
If we do not fix the charge sectors, the level spacing is reduced, and is further reduced if we relax the PHS. However, since the main contributions to the whole spectrum, for $N=9$, is given by $N_f=3,4,5,6,7$ ($89\%$ of the whole spectrum), the deviation from the chaotic behavior in those sectors corresponds approximately to the deviation occurring in the case of the mixed spectrum, either with and without PHS. As a result, we can identify $p_c$, looking at the bending of $r$ from its value at $p=1$, as the point of the overall deviation from chaos in all the charge sectors, as reported in the main text. \\
Once obtained $p_c$ for different values of $N$, we observed that its dependence on the number of sites, for large enough $N$, is consistent with 
\begin{equation}
p_c=\frac{k_cN}{\binom{N}{2}^2}
\end{equation}
(see Fig. \ref{fig:pc}), where $k_c\approx 2.4$. For large $N$, $p_c\simeq \frac{4k_c}{N^3}$. 
A similar behavior is reported in Ref. \cite{preskill-2024}, $p_c=\frac{k_cN}{\binom{N}{4}}$, (for large $N$, $p_c\simeq \frac{24k_c}{N^3}$), with pretty the same coefficient $k_c$, although the number $N_J$ of interactions is different, $N_J=\sum_{i<j<k<l}1=\binom{N}{4}$, since, there, the SYK model describes the interaction among Majorana fermions. In our case, for the complex SYK, the number of interactions is, instead, $N_J=\sum_{i<j, k<l}1=\binom{N}{2}^2$, nevertheless the critical parameter $k_c$ is nearly the same, and, therefore, also the number of remaining interactions, $k_cN$, after the pruning procedure. \\
In order to clarify this behavior, let us review the discussion in Ref. \cite{xu-2020}. In the sparse model, in order to preserve the solvability, namely to have the so-called melon diagrams as leading terms in the $1/N$ expansion,  
we have to normalise the variance of the couplings as $\overline{J_{ijkl}^2}\propto \frac{J^2}{N^3}\frac{1}{p}$. Indeed, after pruning the interactions by $p$, getting, as a result, $\sim pN^3$ residual terms, the basic melon diagram gives a factor proportional to 
\begin{equation}
\frac{J^2}{N^3 p}\times p N^3\propto J^2
\end{equation}
namely, 
the sparsity parameters cancel ensuring that the melon diagram gives the same contribution.
Let us now consider an iterated melon diagram, namely a series of two melon diagrams connected only by a fermionic line. Also in this case, for many configurations of the internal degrees of freedom, these diagrams behave like two copies of melons.
There are, however, special configurations such that the internal indices in a melon diagram are the same, up to a permutation, of the other melon diagram. In this case the couplings should be simultaneously finite (not pruned) with probability $p$. The fraction of these configurations with repeated indices is $\propto 1/N^3$. This kind of diagrams carries two variances of the couplings, $J^4/p^2$ therefore, altogether, it contributes with a factor $\propto{J^4}/{(pN^3)}$. For fixed $p$, e.g. $p=1$, these terms are, therefore, subleading. \\
Let us suppose to have the average number of remaining interactions
to be proportional to $N$, i.e. $L= k N$, while $N_J$ is the number of all possible interactions (or the number of hyperedges in 
a hypergraph with $N$ vertices). The pruning probability, in our case, is, therefore,
\begin{equation}
p=\frac{L}{N_J}=\frac{kN}{\binom{N}{2}^2}\approx \frac{4k}{N^3}
\end{equation}
As a result the subset of iterated melon diagrams, discussed before, contributes with a factor ${J^4}/{(pN^3)}\propto 1/k$. The quantity $k$ plays the role of the degree of a graph. If $k$ grows with $N$, 
the number $L$ of terms grows faster than linearly with $N$ and, we recover the same physics of the fully connected case. 
Also for large but finite $N$ and in the large $k$ limit, $k \gg 1$, we recover the diagrammatics of the fully connected model.
If, instead, we take the large $N$ limit keeping $k$ finite, we still have corrections organised in a $1/k$ expansion. 
Finally, if $k$ decreases with $N$  there is not a controllable hierarchy of the Feynman diagrams and the solvability of the model might be spoiled.\\
In Ref. \cite{preskill-2024}, the authors found that quantum chaos, in the Majorana SYK model, is preserved down to a critical $k$ which they found to be $k_c\approx 2.3$. Quite surprisingly we found almost the same value of $k_c$ for the complex SYK model, although the number of interactions is different.
}

\subsection{Spectral form factor}
The spectral form factor (SFF) provides a robust measure of spectral rigidity 
and it is commonly defined in terms of the analytically continued partition function 
$Z(\tau)$, with $\tau \in \mathbb{C}$,
as follows:
\begin{equation}
\text{SFF}(\beta, t)\equiv\frac{\langle Z(\beta+it)Z(\beta-it)\rangle}{\langle 
Z(\beta)^2\rangle},
\label{sff: definition}
\end{equation}
where $\beta$ is the inverse temperature. 
\textcolor{black}{
Here we consider the partition function $Z(\beta)=Tr[e^{-\beta \hat{H}}]$ where $\beta=1/T$ is the inverse temperature.}

\begin{figure}[h!]
    \centering
    \begin{subfigure}
    \centering
    \includegraphics[width=0.99\linewidth]{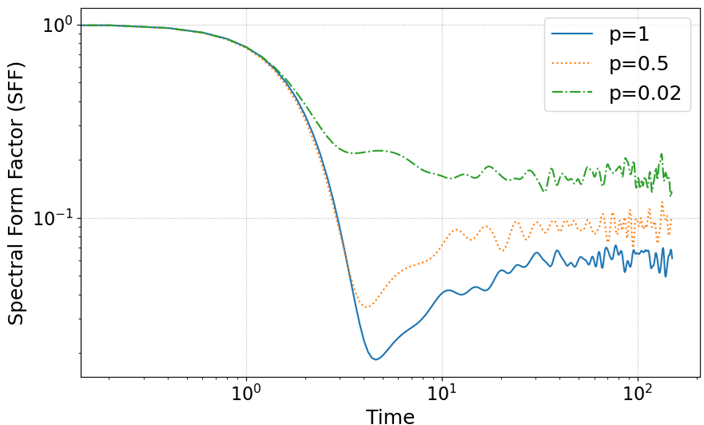}
    \end{subfigure}
    \caption{Spectral form factor (SFF) as defined in Eq. \eqref{sff: definition} 
    vs. time for the SYK model at different sparsity parameters $p$ for $N = 8$. 
    The typical markers of quantum chaos are diluted as $p$ decrease (i.e. we move towards
    to a more sparse scenario), until the dip-ramp feature is completely suppressed around
    $p \simeq 0.02$.
    }
    \label{fig: SFF}
\end{figure}
%

As an exemplary case, we compute the SFF for $N = 8$, 
as reported in Fig. \ref{fig: SFF}.
Evidently, the fully connected case $p=1$ (solid blue line) displays the expected 
dip-ramp-plateau structure characteristic of quantum chaotical systems, with the plateau arising from 
universal long-range spectral correlations. As the sparsity increases (i.e. as $p$ decreases), we 
observe a suppression of the ramp and a lifting of the plateau. This effect becomes more pronounced 
for lower values of the sparsity parameter, such as $p=0.02$ for $N=8$ as displayed in 
Fig. \ref{fig: SFF} (green dashed-dotted line). Technically speaking, the SFF in Fig. \ref{fig: SFF}
is computed in the high-temperature limit, but the quantum chaotical features encoded in the 
energy spectrum are expected to persist across the whole temperature range 
\cite{sonner-2017,saad-2019,preskill-2024}
%
%

\subsection{Population dynamics}
In order to characterize the charging dynamics and
its efficiency, an initial insight can be gained by looking at how energy 
levels are populated through the protocol (i.e. for $t \in [0, \tau_c]$). 
This is indeed provided by the overlap 
\begin{equation}
p_k(t) = \sum_{i} \big|\langle k, i \big| \Psi(t) \rangle\big|^2
\label{population}
\end{equation}
where $\lbrace \big| k,i\rangle \rbrace$ are the eigenstates of $\hat{H}_0= \omega_0 \sum_{i=1}^N \sigma^y_i$, 
the index $i$ labelling the degeneracy related to $\epsilon_k = k \omega_0$, and 
$\Psi(t)$ is the state evolved according to $\hat{H}_B$. Considering
$\hat{H}_0$, 
the initial state is reasonably prepared as
$\big| \Psi_0 \big \rangle = \bigotimes^N_{i=1} \big|\downarrow^{(y)}\big\rangle_i$.
\begin{figure}
    \centering
    \begin{subfigure}
    \centering
    \includegraphics[width=0.99\linewidth]{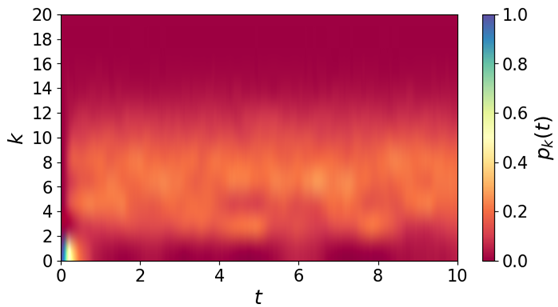}
    \end{subfigure}
    \caption{\textit{Top panel.} 
    Population dynamics $p_k(t)$ as defined in Eq. \eqref{population}
    as a function of time and eigenvalue index $k$; the plot has been obtained 
    for $N = 10$, $J = 2$ and sparsity parameter $p = 0.5$. We have considered here 
    just a single realization of couplings' disorder. Also notice that we have shifted the 
    spectrum upward, such that the $k =  - N$ eigenvalue is labeled by $0$ in the vertical
    axis.
    The peak displayed in the
    bottom-left corner correspond the initial preparation in the ground state
    of $\hat{H}_0$, 
    representing the discharged battery. 
    }
    \label{fig: population dynamics}
\end{figure}
%
%

In Fig. \ref{fig: population dynamics} we report $p_k(t)$ as defined in Eq. \eqref{population} 
for $N = 10$, $J  = 2$, $p = 0.5$ and single realization of couplings' disorder. 
As first noticed in \cite{rossini-2020} regarding the fully connected cSYK model (i.e., the $p = 1$ case), such a charging protocol induces a peculiar non-local dynamics
in energy space. Indeed, as soon as the dynamics unfold for $t > 0$, excited states
are populated almost immediately, singling out a band centered around $k \simeq 8$
persisting at long times. 
This behaviour is intimately related to the peculiar 
properties displayed by the SYK model, 
such as the absence of quasiparticle excitations and the onset of quantum 
chaos, reading local thermalization at the fastest timescale of $\sim \hbar/(k_B T)$ as $T\rightarrow 0$
\cite{hartnoll-2018} ($T$ being the final state temperature), where after-quench 
evolved states can be locally described in terms of thermal ones 
\cite{eberlein-2017,balents-2018}.
%
This is particularly transparent when we 
consider alternative platforms for battery implementation, such as the one 
based on the Dicke model \cite{dicke-1954,kirton-2018,ferraro-2018}, 
where the population dynamics induced by the charging protocol is markedly local,
with evident periodic \textit{revivals}, a feature tracing back to the model's integrability
\cite{dukelsky-2004,strater-2012}.

%
\subsection{Stored energy}
In order to assess the performance of a battery, throughout the charging process, 
it is worth examining specific figures of merit with particular attention to the role
played by the sparsity parameter.
In Fig. \ref{fig: stored energy}, we plot the stored energy, given by
\begin{equation}
E_N(\tau_c) =  \big\langle \Psi(\tau_c) \big| \hat{H}_0 \big| \Psi(\tau_c)\big\rangle - 
E_N^{(0)}
\label{stored energy}
\end{equation}
as a function of $\tau_c$, i.e. the duration of the charging protocol, and 
normalized to the number of cells. The second term in the equation above is simply
the energy of the initial state, namely 
$E_N^{(0)} = \big\langle \Psi_0|\hat{H}_0| \Psi_0\big\rangle$. 
Remarkably, by lowering $p$ (i.e. by increasing sparsity) we see a significant increase in the
energy we are able to stored in the battery. This seemingly points to a trade-off between
the onset of quantum chaos (or rather the ability of the system to be a fast scrambler) and 
the complexity emerging from an all-to-all interacting implementation. A first hint 
supporting this understanding is again reported in Fig. \ref{fig: stored energy}, where 
we see $E_N(\tau_c)$ saturating approximately $p \sim p_c$, a confirmation that the battery 
performance are predicated on retaining at least \textit{some degree} of its quantum-chaotical
behavior in the fully connected ($p = 1$) case.
\begin{figure}[ht!]
    \centering
    \includegraphics[width=0.99\linewidth]{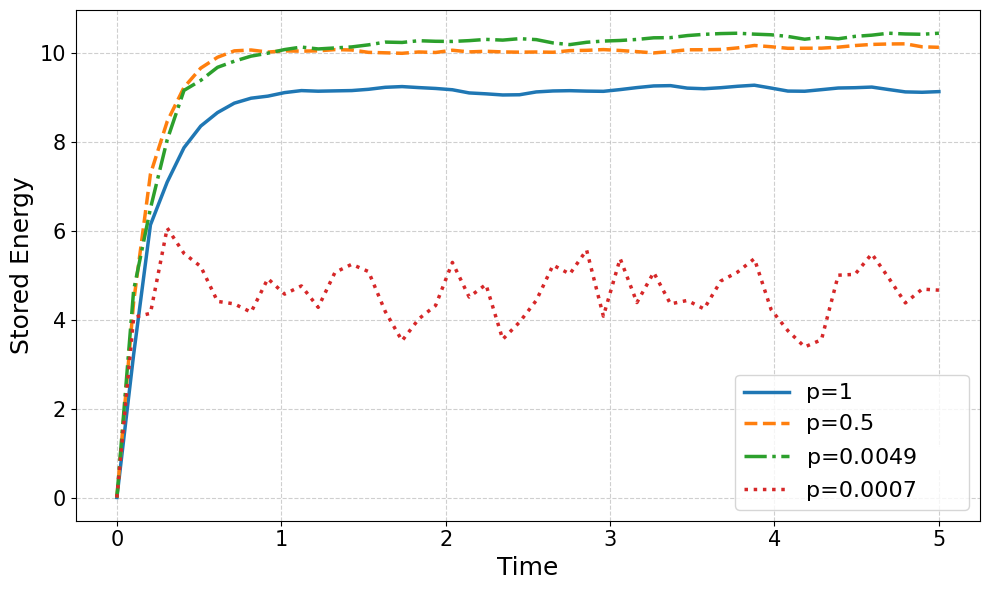}
    \caption{Stored energy (cfr. Eq. \eqref{stored energy}) as a function of $\tau_c$,
    the duration of the charging protocol, and normalized to the number of battery cells.
    Here we consider $N = 12$, $J = 3$ and four different values of the sparsity parameter.
    Additionally, $E_N(\tau_c)/N$ is also averaged over $N_{\text{dis}} = 50$ different 
    disorder realization of the model's couplings $\lbrace J_{ijkl} \rbrace$.
    The fully-connected cSYK ($p = 1$) is reported as a solid blue, while dash-dotted green line
    follows the case of $p_c \simeq 0.0049$ (for $N = 12$). 
    }
    \label{fig: stored energy}
\end{figure}

\subsection{\textcolor{black}{Efficiency}}
\begin{figure}[ht!]
    \centering
    \includegraphics[width=0.99\linewidth]{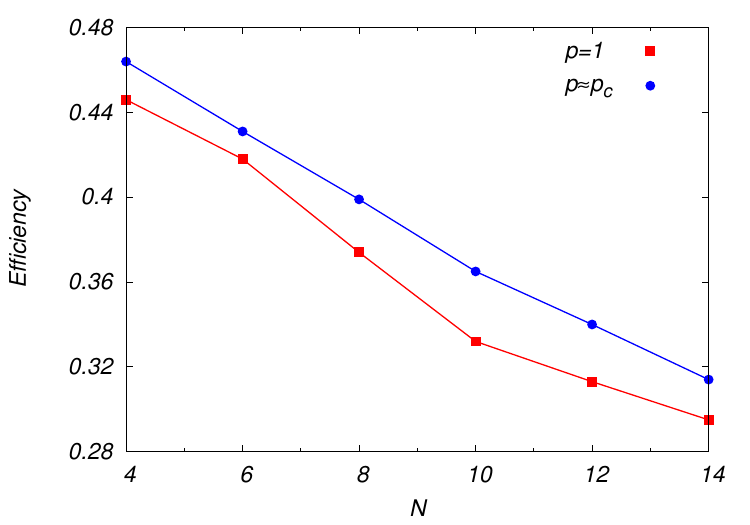}
 \caption{\textcolor{black}{Scaling with the system size $N$ of the efficiency, Eq. (\ref{efficiencyM}), for $M=N/2$, with the same parameters of Fig. 4 in the main text. We report the values of the efficiency, for $p=1$ (red squares) and its maximum values obtained by pruning the interactions 
 for $p\lesssim p_c$ (blue dots).}}
    \label{fig:efficiency_N}
\end{figure}

\textcolor{black}{Let us consider the efficiency defined as the ratio between the amount of extractable work and the energy stored in the battery during the charging process 
\begin{equation}
e(\tau_c) = \overline{\bigg(\frac{\mathcal{E}_{M}(\tau_c)}{E_{M}(\tau_c)}\bigg)}
\label{efficiencyM}
\end{equation}
 where ${\cal E}_{M}$ and  ${E}_{M}$ are the ergotropy and the stored energy in a portion of the system composed by $N$ sites,  with $M\le N$, of course, for $M=N$, $e(\tau_c) =1$. The profile of  $e(\tau_c) $, as a function of $M/N$, has been used to detect many-body localized, Anderson localized, and ergodic phases in a quantum battery described by a disordered quantum Ising chain \cite{rossini-2019}. 
 In Fig. 4 of the main text, where $M=N/2$, we see that the efficiency is improved by pruning the interactions, when approaching the critical sparsity parameter $p_c$, gaining up to $10\%$, at least for the parameters used in that figure. 
 We see, however, that the efficiency of the battery, when considering half of the system, decreases as the number $N$ increases, see Fig. \ref{fig:efficiency_N}. This result suggests that the 
 the stored energy increases faster than the extractable energy as the thermodynamic limit is approached. 
 This observation is in line with a similar result reported in \cite{rosa-2020}  where the authors examined the fraction of energy per cell, concluding that considering very large batteries using SYK protocol is not so beneficial. 
From Fig. \ref{fig:efficiency_N}  we cannot argue the profile of the decrease, also because of objective computational limits (currently, for the same set-up, the largest system size reached is $N=16$), however we can clearly see that the trend of the efficiency is approximately the same for the fully connected model and the sparse one, with a curve always above the other. 
Pruning the interactions seems, therefore, to promote the efficiency of the battery for any size of the system. 
However, more than the scaling of the efficiency, the key figure of merit useful to directly access the quantum advantage, the superior performance that quantum systems can exhibit over their classical counterparts, is the scaling of the charging power. 
 }

\subsection{\textcolor{black}{Charging power}}

\begin{figure}
     	\includegraphics[height=4.25cm,width=4.25cm]{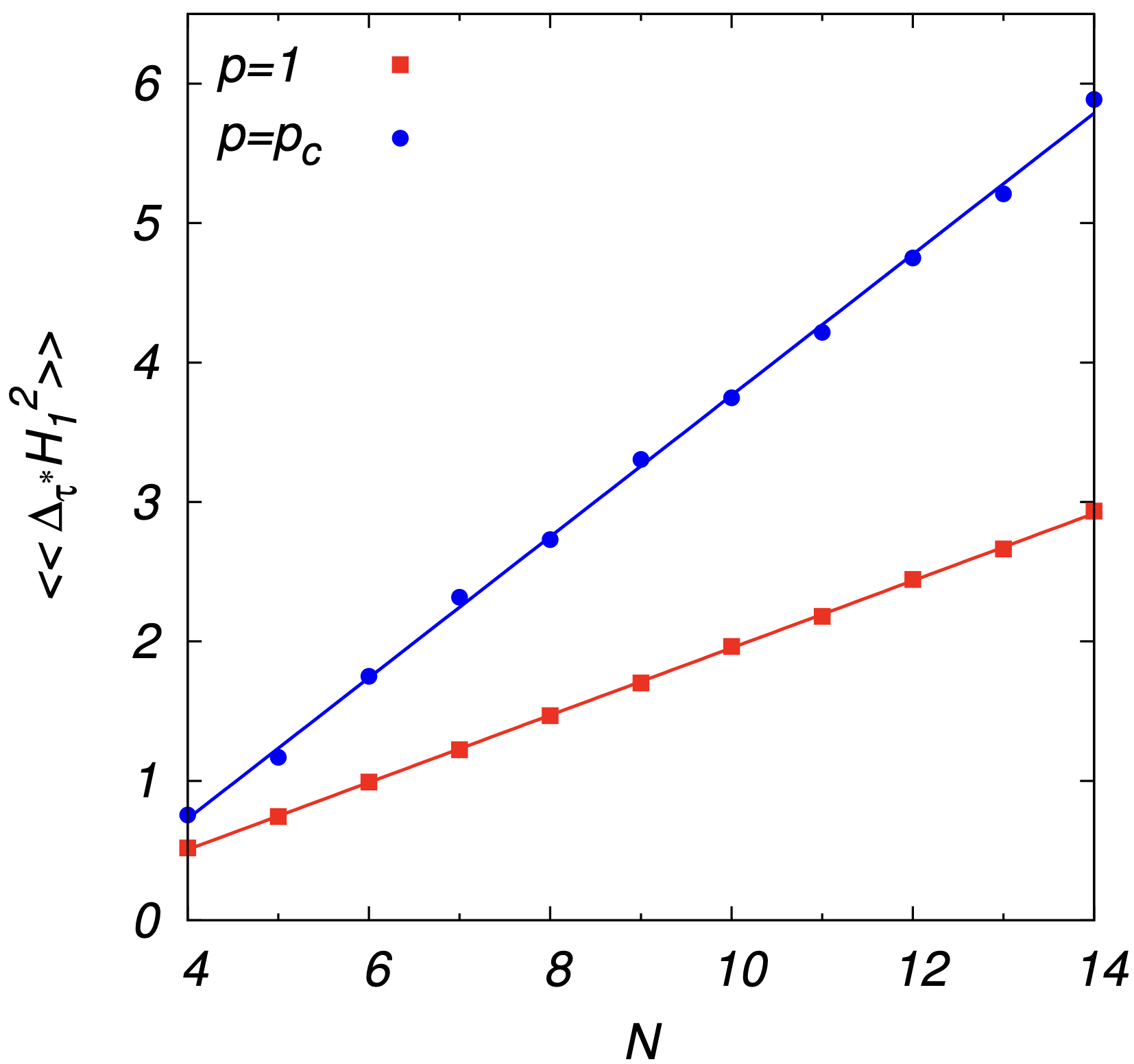}
         \includegraphics[height=4.25cm,width=4.25cm]{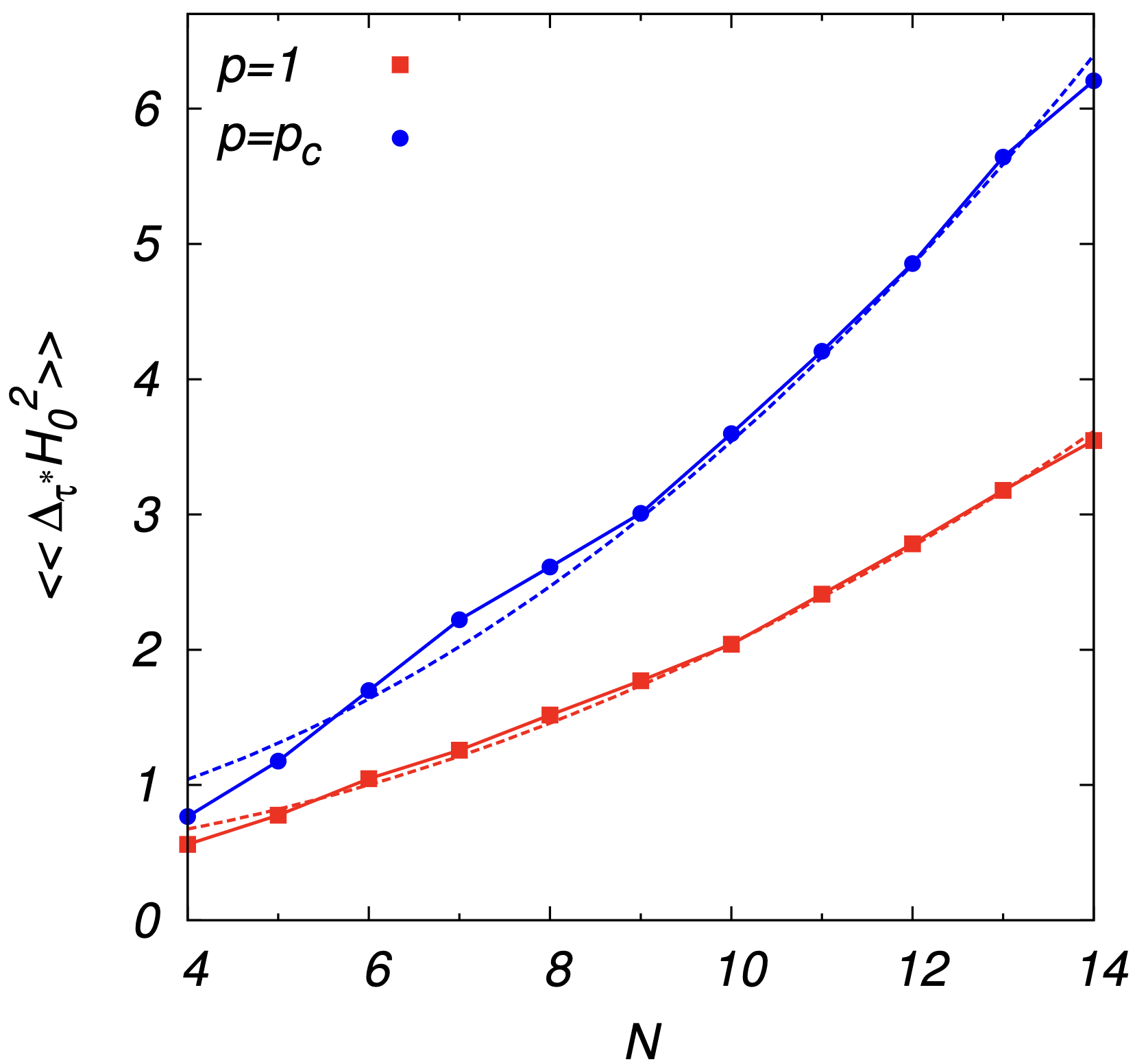}\\
          \includegraphics[height=4.25cm,width=4.25cm]{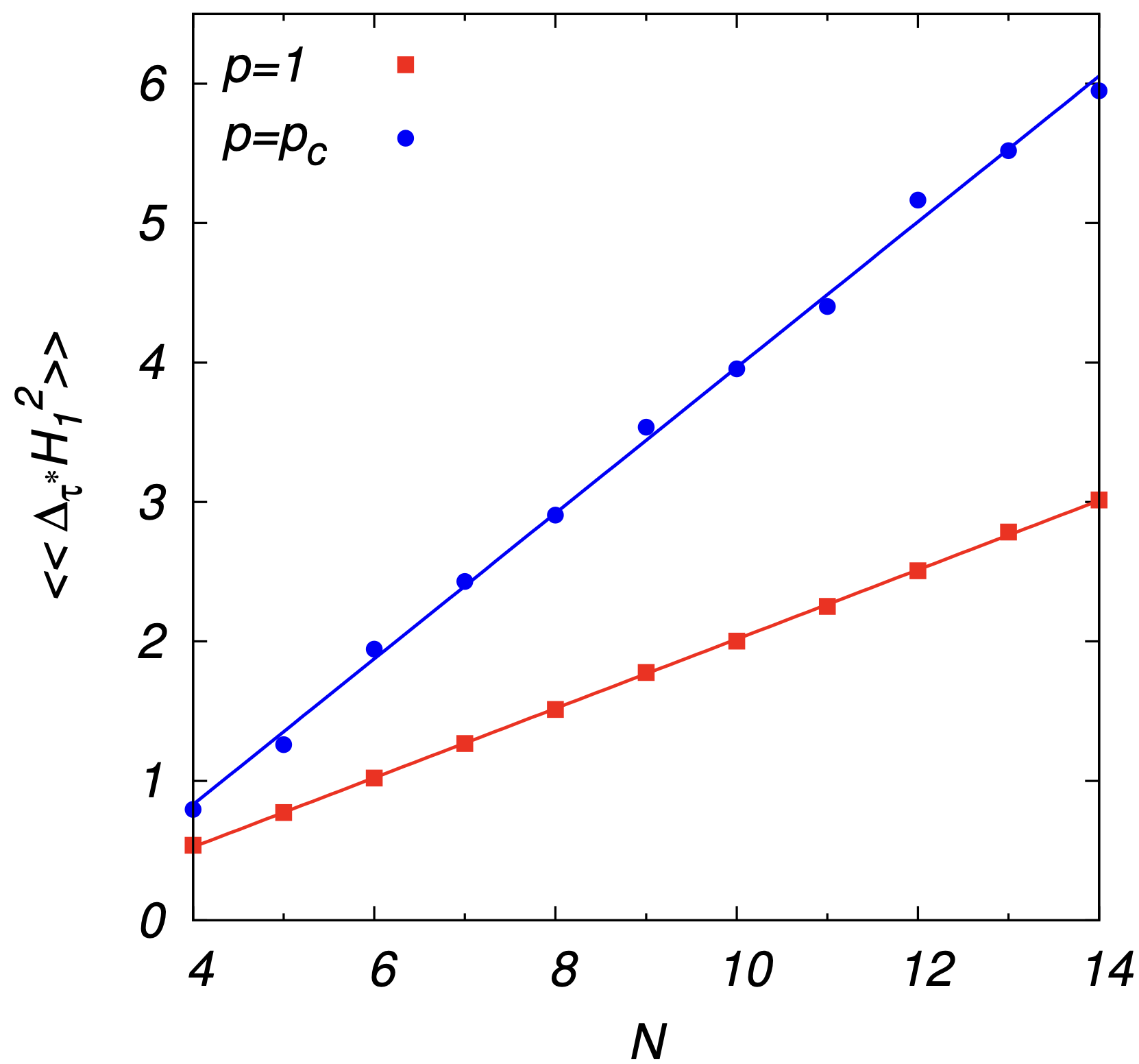}
          \includegraphics[height=4.25cm,width=4.25cm]{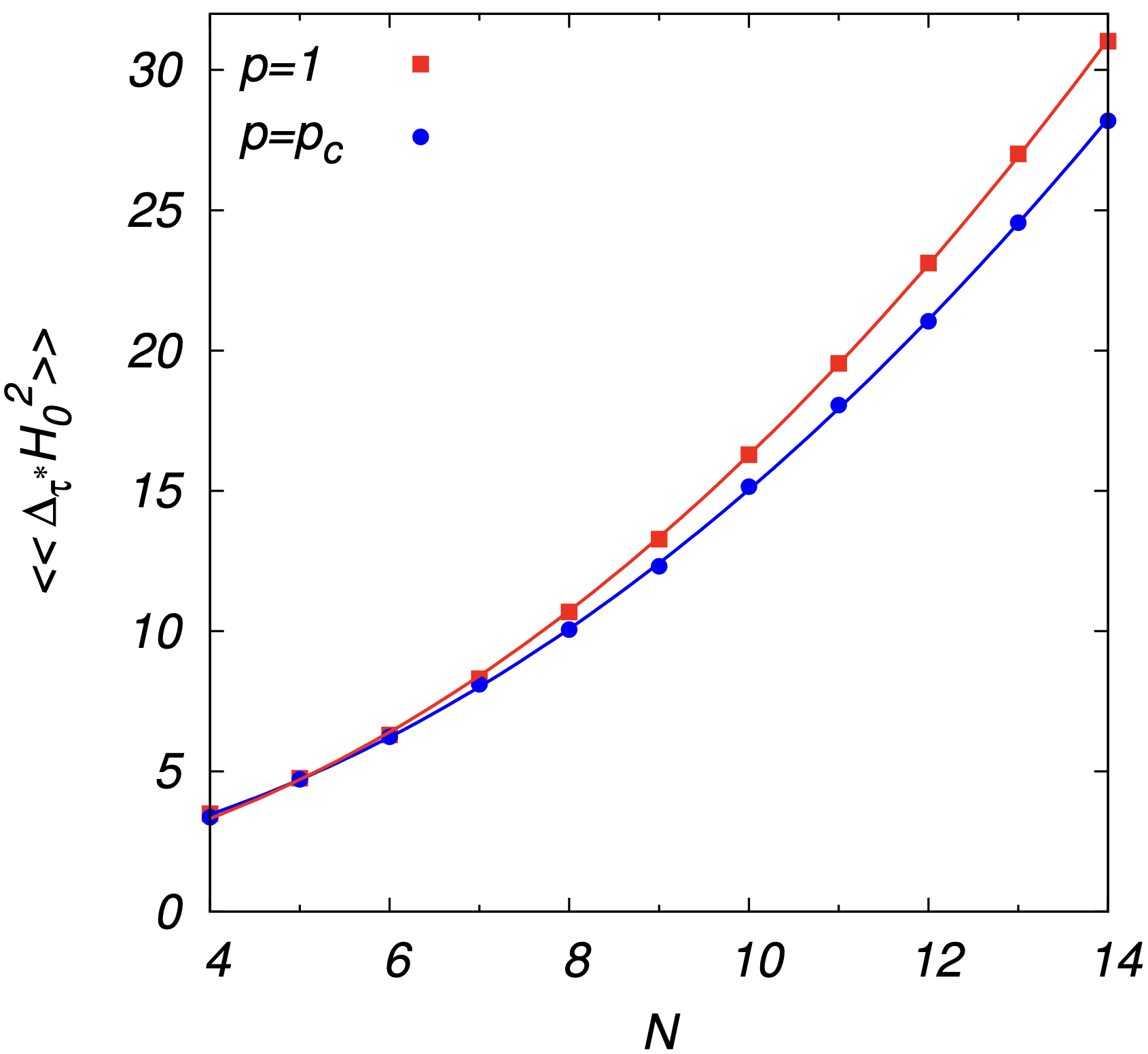}
    \caption{
    \textcolor{black}{
    Scaling with the system size $N$ of the quantities 
    $\langle\langle \Delta_{\tau^\star} \hat H_1^2 \rangle\rangle$, in units of $J^2$ (left panels), and $\langle\langle \Delta_{\tau^\star} \hat H_0^2 \rangle\rangle$, in units of $\omega_0^2$ (right panels), for two different sparsity regimes: 
    the red squares correspond to the fully connected case ($p=1$), while the blue circles refer to the sparse regime ($p=p_c$), 
    calculated for $J=1$ (upper panels) and $J=100$ (lower panels)
    The solid lines in the upper left, lower left and lower right panels are the best fits with linear ($\sim N$, left panels) and quadratic ($\sim N^2$, lower right panel) functions. In the upper right panel the quadratic behaviors is less clear (and displayed by dashed lines while the solid line is a guide to the eye.
    }}
    \label{fig:Advantage_HO2}
\end{figure}

\begin{figure}
     	\includegraphics[width=0.99\linewidth]{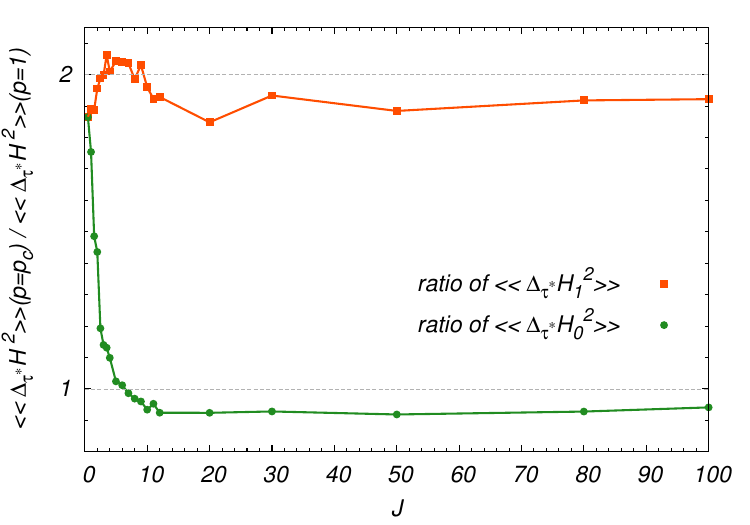}
    \caption{\textcolor{black}{Ratios between $\langle\langle \Delta_{\tau^\star} \hat H_\alpha^2 \rangle\rangle$ for $p=p_c$ and the same quantities calculated for $p=1$, as functions of the strength of the interaction $J$, for $N=8$. The orange squares represents ${\langle\langle \Delta_{\tau^\star} \hat H_1^2 \rangle\rangle(p=p_c)}/{\langle\langle \Delta_{\tau^\star} \hat H_1^2 \rangle\rangle(p=1)}$ which is almost $2$ for any values of $J$, while ${\langle\langle \Delta_{\tau^\star} \hat H_0^2 \rangle\rangle(p=p_c)}/{\langle\langle \Delta_{\tau^\star} \hat H_0^2 \rangle\rangle(p=1)}$, represented by the green circles, is around $1$, for $J\gtrsim 3$.}}
    \label{fig:N8}
\end{figure}

\begin{figure}
     	\includegraphics[width=0.99\linewidth]{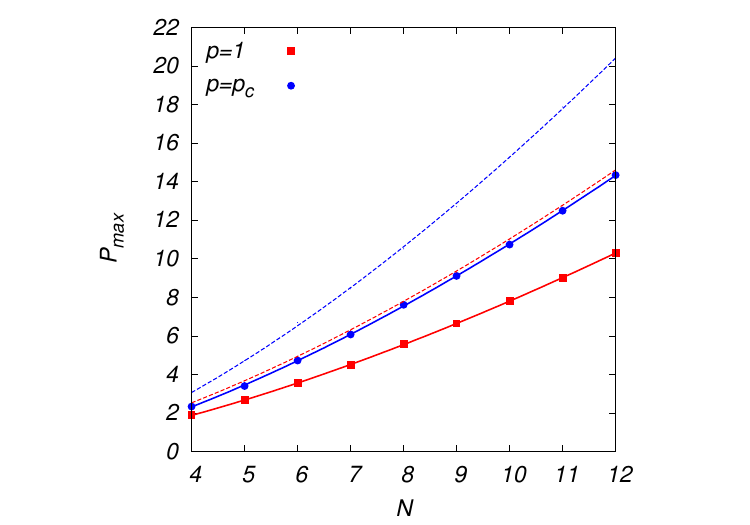}
\caption{\textcolor{black}{$P_{max}$, in units of $J\omega_0$, for $p=1$ (red squares) and $p=p_c$ (blue dots), fitted by $\alpha_1 N^{3/2}+\beta_1$  with $\alpha_1\approx 0.25$, $\beta_1\approx -0.13$ (red solid line) and $\alpha_c N^{3/2}+\beta_c$ with $\alpha_c\approx 0.36$, $\beta_c\approx -0.54$, (blue solid line). The dashed lines represent the upper bounds $2\sqrt{\Delta_{\tau^*} \hat{H}^2_0 \, \Delta_{\tau^*} \hat{H}^2_1}$, for $p=1$ (red dashed line) and $p=p_c$ (blue dashed line).}}
   \label{fig:Pmax}
\end{figure}

 \textcolor{black}{The charging power is one of the main features useful to check the quantum performance of the quantum batteries. As shown in the main text its value is bonded by the energy fluctuations,
 \begin{equation}
 P(\tau)\le 2\sqrt{\Delta_\tau \hat{H}^2_0 \, \Delta_\tau \hat{H}^2_1}
 \label{app.Power}
\end{equation}
This relation comes from the Heisenberg equation of motion, 
$(\frac{dE_N(t)}{dt})^2=|\langle[\hat H_0,\hat H_1]\rangle_t|^2$, for $0\le t\le \tau$, and the Heisenberg-Robertson-Schr\"odinger uncertainty relation $|\langle[\hat H_0,\hat H_1]\rangle_t|^2\le 4\,\delta_t \hat H_0^2\, \delta_t \hat H_0^2$, with $\delta_t \hat H^2=\langle\hat H^2\rangle_t-\langle \hat H\rangle_t^2$. Taking the square root and the time average, $\int \frac{dt}{\tau}$, fixing $E_N(0)=0$ so that $P(\tau)=E_N(\tau)/\tau$, and finally using the Cauchy-Schwarz inequality
$\int \frac{dt}{\tau} \sqrt{\delta_t \hat H^2_0}\sqrt{\delta_t \hat H^2_1}\le \sqrt{\int \frac{dt}{\tau}  \delta_t \hat H^2_0}\sqrt{\int\frac{dt}{\tau}  \delta_t \hat H^2_1}$, one gets Eq. (\ref{app.Power}).\\
The results displayed in Fig.~5 of the main text illustrate the scaling of $\langle\langle \Delta_{\tau^\star} \hat H_{\alpha}^2 \rangle\rangle$, namely $\Delta_{\tau^\star} \hat H_{\alpha}^2$ averaged over disorder \cite{rossini-2020}, with $N$, evaluated at the optimal time $\tau^\star$, for which $P(\tau^\star)=P_{max}$, for two distinct connectivity regimes: the fully connected case ($p=1$) and the sparse regime ($p=p_c$).  Those plots are calculated for $J=10$. Here, in Fig.~\ref{fig:Advantage_HO2} the same quantities are calculated for $J=1$ (upper panels) and $J=100$ (lower panels).  For the charging power we choose to work preserving PHS in order to better compare our results with those reported in \cite{rossini-2020}. However, we check that, as already said, there are no qualitative differences if we do not impose PHS. In the latter case the maximum values of the charging power are slightly higher.  \\
We observe that $\langle\langle \Delta_{\tau^\star} \hat H^2_1 \rangle\rangle$, in units of $J$, both for $p=1$ and $p=p_c$, is almost always the same, for any value of $J$ (see the upper left panel of Fig.~\ref{fig:Advantage_HO2}, where $J=1$,  the left panel of Fig.~5 in the main text, where $J=10$, and the lower left panel of Fig.~\ref{fig:Advantage_HO2}, where $J=100$). 
On the contrary, $\langle\langle \Delta_{\tau^\star} \hat H^2_0 \rangle\rangle$ for $p=p_c$ is much greater than that for $p=1$, for small $J$ (see the upper right panel of Fig. ~\ref{fig:Advantage_HO2}) while it takes almost the same values for large enough $J$ (see the right panel of Fig.~5 in the main text, where $J=10$, and the lower right panel of Fig. ~\ref{fig:Advantage_HO2}, where $J=100$). \\
These behaviors are highlighted in Fig.~\ref{fig:N8}, where the ratios $\frac{\langle\langle \Delta_{\tau^\star} \hat H_1^2\rangle\rangle(p=p_c)}{\langle\langle \Delta_{\tau^\star} \hat H_1^2\rangle\rangle (p=1)}$ (orange line) and $\frac{\langle\langle \Delta_{\tau^\star} \hat H_1^2\rangle\rangle(p=p_c)}{\langle\langle \Delta_{\tau^\star} \hat H_1^2\rangle\rangle (p=1)}$ (green line)
were calculated at fixed $N$ ($N=8$) as functions of $J$. One can clearly see that, for any values of $J$,
\begin{equation}
\langle\langle\Delta_{\tau^*} \hat{H}^2_1\rangle\rangle (p=p_c)\simeq 2 \langle\langle\Delta_{\tau^*} \hat{H}^2_1\rangle\rangle (p=1)
\end{equation}
while, for $J\gtrsim 3$, 
\begin{equation}
\langle\langle\Delta_{\tau^*} \hat{H}^2_0\rangle\rangle  (p=p_c)\simeq  \langle\langle\Delta_{\tau^*} \hat{H}^2_0\rangle\rangle (p=1)
\end{equation}
Actually the latter quantity, approaches very soon the ratio ($c'/c$) of the factors of the quadratic behaviors obtained by fitting the plots in the right panel of Fig.~5 
of the main text, or analogously, the plots in the lower right panel of Fig.~\ref{fig:Advantage_HO2}, 
\begin{eqnarray}
&&\frac{1}{\omega_0^2}\langle\langle\Delta_{\tau^*} \hat{H}^2_0\rangle\rangle (p=1)\approx c \,N^2+d\\
&&\frac{1}{\omega_0^2}\langle\langle\Delta_{\tau^*} \hat{H}^2_0\rangle\rangle (p=p_c)\approx c' N^2+d'. 
\end{eqnarray}
The best fits give $c\approx 0.15$ ($d\approx 0.72$) and $c'\approx 0.14$ ($d'\approx 0.94$), therefore $c'/c\approx 0.93$. 
For the fluctuations related to the charging Hamiltonian we calculated the best linear fits, 
\begin{eqnarray}
&&\frac{1}{J^2}\langle\langle\Delta_{\tau^*} \hat{H}^2_1\rangle\rangle (p=1)\approx a\,N+b \\
&&\frac{1}{J^2}\langle\langle\Delta_{\tau^*} \hat{H}^2_1\rangle\rangle(p=p_c)\approx a' N+b' 
\end{eqnarray}
with $a\approx 0.23$ ($b\approx -0.40$) and $a'\approx 0.52$ ($b'\approx -1.36$). For very large $N$ we expect, therefore, that the ratio of these two quantities approaches $a'/a\approx 2.25$. 
Using Eq. (\ref{app.Power}) at $\tau=\tau^*$, we expect, therefore,
\begin{equation}
\frac{P_{max}(p=p_c)}{P_{max}(p=1)}\approx \sqrt{\frac{a'c'}{a\, c}}\approx \sqrt{2}.
\end{equation}
Actually this ratio is confirmed by numerically calculating $P_{max}$, for $p=1$ and $p=p_c$. As shown in Fig. \ref{fig:Pmax} 
the maximum charging powers, are very well fitted by
\begin{eqnarray}
&&\frac{1}{J\omega_0} P_{max} (p=1)\approx \alpha_1 N^{3/2}+\beta_1 \\
&&\frac{1}{J\omega_0} P_{max} (p=p_c)\approx \alpha_c N^{3/2}+\beta_c 
\end{eqnarray}
with the best parameters reported in Fig. \ref{fig:Pmax}. In particular, we obtain that $\frac{P_{max}(p=p_c)}{P_{max}(p=1)}\approx \frac{\alpha_c}{\alpha_1}\approx \sqrt{2}$, validating what we expect analysing the upper bounds. \\
Intriguingly we observe, at the end, that the upper bound $2\sqrt{\Delta_{\tau^*} \hat{H}^2_0 \, \Delta_{\tau^*} \hat{H}^2_1}$, in the fully connected case ($p=1$) (red dashed line in Fig. \ref{fig:Pmax}) almost coincides with $P_{max}$ in the critically sparse case ($p=p_c$) (blue solid line in Fig. \ref{fig:Pmax}).
}

\subsection{Additional experimental considerations}
Starting from a single-mode
optical cavity filled with quasi-2D $^6$Li atoms, through fast cycling and engineered time-dependent 
disorder, it is possible to simulate the evolution of a random quantum circuit through discrete
steps, converging to the desired model. Remarkably, this hybrid digital-analog technique (resembling conventional Trotterization) 
is able to target the sparse version of the desired target model (the full SYK, in this case) \cite{hauke-2024}.
In order to estimate sparseness, one can compute the \textit{statistical distance} from the fully connected
model via the so-called Kullback-Leibler divergence (or relative Shannon entropy), namely 
$D_{KL}(\mathcal{J}||\mathcal{J}_{sp}) = \int \mathcal{D}[\mathcal{J}]\mathcal{P}(\mathcal{J})
\log(\mathcal{P}(\mathcal{J})/\mathcal{Q}(\mathcal{J}_{sp}))$ where 
$\mathcal{P}(\mathcal{J} = J_{ijkl})$ is the distribution of the fully-connected model while 
$\mathcal{Q}(\mathcal{J}_{sp})$ describess the sparse one.
 
\end{document}